\documentclass[aps,prb,twocolumn,superscriptaddress,
floatfix,notitlepage]{revtex4-1}

\usepackage{amsthm}

\usepackage{graphicx,psfrag,color}
\usepackage{mathbbol,amsmath,amsfonts,amssymb,bm,dsfont,braket}
\usepackage{wasysym}
\usepackage[section]{placeins}
\usepackage{multirow}

\def\x{{\mathbf{x}}}
\def\y{{\mathbf{y}}}

\newcommand{\ignore}[1]{ }

\usepackage{verbatim}

\usepackage{array}
\newcolumntype{x}[1]{>{\centering\hspace{0pt}}p{#1}}
\usepackage[normalem]{ulem}

\begin{document}

\title{Bloch and Bethe Ans\"{a}tze for the Harper model: A butterfly with a boundary}

\begin{abstract}
Based on a recent generalization of Bloch's theorem, we present a Bloch Ansatz for the Harper model with an arbitrary rational magnetic flux in various geometries, and solve the associated Ansatz equations analytically. In the case of a cylinder and a particular boundary condition, the energy spectrum of edge states has no dependence on the length of the cylinder, which allows us to construct a quasi-one-dimensional edge theory that is exact and describes two edges simultaneously. We prove that energies of bulk states, generating the so-called Hofstadter's butterfly,  depend on a single geometry-dependent spectral parameter and have exactly the same functional form for the cylinder and the torus with general twisted boundary conditions, and argue that the (edge) bulk spectrum of a semi-infinite cylinder in an irrational magnetic field is (the complement of) a Cantor set. Finally, realizing that the bulk projection of the Harper Hamiltonian is a linear form over a deformed Weyl algebra, we introduce a Bethe Ansatz valid for both cylinder and torus geometries.
\end{abstract}

\author{Qiao-Ru Xu}
\affiliation{\mbox{Department of Physics, Indiana University, Bloomington, Indiana 47405, USA}}
\author{Emilio Cobanera} 
\affiliation{\mbox{Department of Mathematics and Physics, SUNY Polytechnic Institute, 100 Seymour Rd, Utica, New York 13502, USA }}
\affiliation{\mbox{Department of Physics and Astronomy, Dartmouth College, 6127 Wilder Laboratory, Hanover, New Hampshire 03755, USA}} 
\author{Gerardo Ortiz}
\affiliation{\mbox{Department of Physics, Indiana University, Bloomington, Indiana 47405, USA}}
\affiliation{Indiana University Quantum Science and Engineering Center, Bloomington, Indiana 47408, USA}

\date{\today}
\maketitle


\section{Introduction} 
In hindsight, the discovery of the integer quantum Hall effect \cite{Klitzing} in 1980  
started one of the most active subfields of condensed matter physics, that of 
electronic topological quantum matter \cite{Chiu}. The effect can be understood  
in terms of independent electrons moving on a plane, or on a  planar periodic lattice, subject
to a homogeneous magnetic field of magnitude \(B\) applied in the direction perpendicular to 
the plane. The case of a square (or cubic) lattice was investigated by Harper \cite{Harper} and later on 
by Azbel' \cite{Azbel} and Hofstadter \cite{Hofstadter}. The Harper model has played an important role 
at various stages of the development of the theory of topological quantum  matter. 
Its infinitely many quantum phases (in the flux per plaquette versus filling fraction diagram) 
are labeled by the TKNN topological invariant \cite{TKNN}, a Chern number. Most importantly,  
the Harper model defined on a manifold with boundaries displays topologically-protected 
boundary states, a manifestation of the so-called bulk-boundary correspondence \cite{Hatsugai}. 

Another fascinating aspect of the Harper model, first pointed out by Hofstadter \cite{Hofstadter}, is 
that it  generates a Cantor (multifractal)  set when defined in a planar periodic lattice
with no boundaries. The set in question, known as 
``Hofstadter's butterfly'' for its visual representation, is the collection of energy levels of the Harper 
Hamiltonian for a range of values of the flux per square plaquette. By contrast, the energy spectrum 
of the continuous problem, the Landau levels, is identical to that of a harmonic oscillator
of frequency \(\omega_c=|e|B/{\sf m}\), the cyclotron frequency (in this paper, \(-e\) and \(\sf m\) 
are the charge and mass of the particles).  How can the presence of a periodic lattice change the spectral 
complexity of the problem so dramatically? The answer is that, in the continuum \cite{Ballentine}, the electrons
move in quantized circular orbits of radius $r_c$, with \(r_c^2=\frac{\hbar}{{\sf m}\omega_c}(2{\sf n}+1)\) 
and \(\sf n\) a non-negative integer. Hence, the flux enclosed by the path of the electron in an energy 
eigenstate is \(\Phi_{\sf n}=B\pi r^2_c=({\sf n}+\frac{1}{2})\phi_0\) and is always commensurate with 
the flux quantum \(\phi_0=h/|e|\). By contrast\cite{Fradkin}, a lattice introduces an additional length scale and 
the electron can follow a path enclosing an irrational value of magnetic flux in units of \(\phi_0\).

The experimental realization of Hofstadter's butterfly has a colorful history. Starting with the transmission 
of microwaves through an array of scatterers \cite{Kuhl}, to semiconducting quantum Hall\cite{Albrecht} or 
graphene devices\cite{Ponomarenko,Dean,Hunt}, it has long been a subject of admiration perhaps as a 
result of its beauty and deep connections to diverse mathematical concepts, such as number theory and 
Apollonian gaskets \cite{Satija}. It was not until recent years that the simplest square-lattice Harper 
Hamiltonian was realized in optical lattices loaded with ultracold bosonic rubidium ($^{87}\text{Rb}$) 
atoms \cite{Bloch,Ketterle}. Chiral edge states of Harper Hamiltonians implemented in optical lattices 
were also observed, using either ultracold fermionic ytterbium ($^{173}\text{Yb}$) atoms \cite{Zoller} 
or bosonic rubidium ($^{87}\text{Rb}$) atoms \cite{Spielman}. 

We notice that the latter two experiments 
\cite{Zoller,Spielman} are equipped with long strip lattices that can be well approximated by the thin 
cylinder geometry. Is it possible to get complete analytical solutions, both for the bulk and the boundary, 
in the case of a finite-sized or semi-infinite cylindrical Harper system? If so, what important 
information, not possible to obtain from numerics, can be extracted from these analytical solutions? 
For example, does the Cantor set nature of the bulk spectrum persist for a system with boundaries? After 
all, experiments are performed in samples with boundaries. What is 
the effective boundary theory of the Harper model and how does it evolve when the boundary changes? 
What is the topological character of the boundary theory and does it carry information about the fractal 
nature of the bulk spectrum? In this paper, we address these questions\cite{Comment} and elaborate on 
some other curiosities based on an analytical tool dubbed generalized Bloch theorem \cite{PRL1,BlochAnsatz,PRB1,PRB2}.

The outline of the paper is as follows. In Sec.\,\ref{HarperHofstadterHamiltonians}, we describe 
the Harper model in some detail for torus and cylinder geometries and describe its notable symmetries in 
real space (Appendix\,\ref{duality} is on duality transformations that map the Harper model to superconductors). 
In Sec.\,\ref{BetheAnsatzSolutions}, we establish the exact Bloch Ansatz single-particle 
wavefunctions, together with the single-particle energies, for the cylinder geometry (technical details for 
exceptional bulk solutions are presented in Appendix\,\ref{degenerateroots}). Since the Harper model is 
number-conserving, our Ans\"{a}tze are equally useful for modeling both fermions and bosons of the 
putative many-body problem \cite{FermionBoson}. It has been rigorously shown \cite{Math} that the 
spectrum of the Harper model defined in an infinite plane forms a Cantor set in the case of an irrational flux. 
Here, we argue, based on embeddings of bulk spectra, that the same Cantor set structure is preserved in the 
case of the cylinder (the semi-infinite case is analytically solved in Appendix\,\ref{semi-infinite}, and the general 
case of twisted boundary conditions is presented in Appendix\,\ref{TwistedBCs}). Furthermore, we construct an 
exact boundary theory from boundary solutions (including two edges) which happens to be topologically trivial 
independently of the value of the flux per square plaquette (Appendix\,\ref{nuneq1} studies the effect of certain 
boundary conditions on the boundary theory). In Sec.\,\ref{Berus}, we introduce a deformation of the 
Weyl algebra (the connection to the Yang-Baxter equation is in Appendix\,\ref{YBE}) and show that the bulk 
Harper Hamiltonian for the cylinder is a linear form over generators of this algebra. We then present an algebraic 
Bethe Ansatz in some key aspects simpler than the one developed in Ref.\,[\onlinecite{Wiegmann}]. Our Bethe 
Ansatz is applicable to both the (bulk projected) Hamiltonian defined on a cylinder or torus, the difference being 
encoded in the specific functional form of a spectral parameter. Section\,\ref{compare} is dedicated to comparing 
our Ans\"{a}tze to the Bethe Ansatz of Ref.\,[\onlinecite{Wiegmann}]. We conclude in Sec.\,\ref{Summary}.

\section{The Harper model }
\label{HarperHofstadterHamiltonians}

\begin{figure}[t]
\includegraphics[width=\columnwidth]{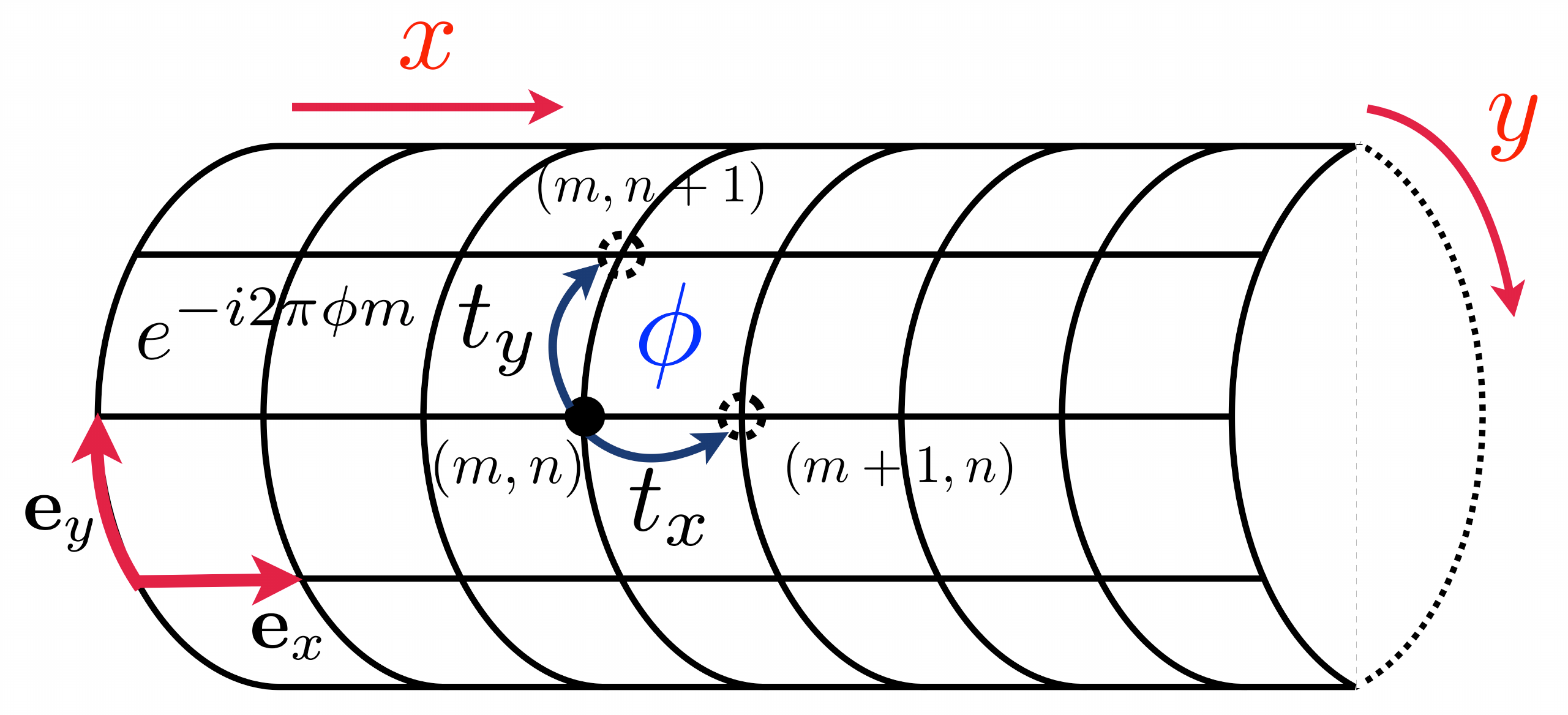}
\caption{The cylinder geometry of the Harper model, where $\phi$ is the magnetic flux per plaquette (in units of the flux quantum $\phi_0$). 
A fermion or a boson of charge $-e$ at site $(m,n)$, with the nearest-neighbor hopping amplitude in the $x\,(y)$ direction being $t_x\,(t_y)$, 
picks up a phase $e^{-i2\pi\phi m}$ when it hops along $\mathbf{e}_y$ to site $(m,n+1)$ and picks up a total phase $e^{-i2\pi\phi}$ when it 
hops counterclockwise around a plaquette.}
\label{fig:Cyl}
\end{figure}

The second-quantized Hamiltonian of spinless particles confined 
to a square lattice in the $xy$-plane is
\[
\widehat{H}=
\sum_{m,n}\left( t_x c_{m,n}^\dagger c^{\;}_{m+1,n}+t_y c_{m,n}^\dagger c^{\;}_{m,n+1} +\text{H.c.}\right).
\]
The lattice sites are labeled by an ordered pair of integers \((m,n)\). We will not specify the boundary conditions right away. 
The $c_{m,n}^\dagger\,(c^{\;}_{m,n})$ are the creation (annihilation) operators of fermions or bosons
of charge $-e$. With the exception of the duality transformations, our work in this paper is insensitive to the quantum
statistics of the particles and \(-e\) is not necessarily the charge of the electron but rather the charge variable. When the system is coupled to a slowly varying magnetic vector potential $\mathbf{A}$, one can use the prescription known as the Peierls substitution to revise the above Hamiltonian as follows
\begin{align}
\widehat{H}=&\sum_{m,n}\left(t_xe^{i\frac{-e}{\hbar}\int_{(m+1,n)}^{(m,n)}\mathbf{A}\cdot d\mathbf{r}}c_{m,n}^\dagger c^{\;}_{m+1,n}+\text{H.c.}\right)\nonumber\\
+&\sum_{m,n}\left(t_ye^{i\frac{-e}{\hbar}\int_{(m,n+1)}^{(m,n)}\mathbf{A}\cdot d\mathbf{r}}c_{m,n}^\dagger c^{\;}_{m,n+1} +\text{H.c.}\right).\nonumber
\end{align}
The Harper model \cite{Harper} is the special case 
when $\mathbf{A}$ is that of a uniform magnetic field pointing in the \(z\) direction perpendicular to the lattice. Then, one can choose to work in the Landau gauge $\mathbf{A}=\frac{h}{e}\phi x\mathbf{e}_y$ so that
\begin{align}\label{Harper-Hofstadter}
\hspace*{-0.2cm}\widehat{H}\hspace*{-0.05cm}=\hspace*{-0.05cm}
\sum_{m,n}\big( t_x c_{m,n}^\dagger c^{\;}_{m+1,n}+t_y e^{i2\pi\phi m} c_{m,n}^\dagger c^{\;}_{m,n+1} +\text{H.c.}\big),\hspace*{-0.2cm}
\end{align}
and $\phi$ is the magnetic flux per plaquette, measured in units of the flux quantum $\phi_0=h/|e|$ where $h$ is Planck's constant  (see Fig.\,\ref{fig:Cyl}). For electrons, \(\phi_0\approx 4.1\times 10^{-15}\,\text{Wb}\). 

\subsection{Boundary conditions} 

If the magnetic flux per plaquette is rational, i.e., $\phi=P/Q$ with $P$ and $Q$ co-prime integers, then the Harper Hamiltonian of  Eq.\,(\ref{Harper-Hofstadter}) becomes translation invariant. The magnetic unit cell contains $Q$ neighboring lattice sites on a line in the $x$ direction because
\[
e^{i2\pi \phi m}=e^{i2\pi \phi (j_Q+s)}=e^{i2\pi \phi s}=\omega^{s},
\]
where $j_Q=Q(j-1)$ with \(j\) an integer, 
 $s\in\{1,2,\dots,Q\}$, and $\omega=e^{i2\pi \phi }$. So, translations by \(Q\) sites in the $x$ direction are symmetries. It is convenient to decompose the subspace spanned by the single-particle orbital states 
$|m,n\rangle$ 
as the tensor product $| j \rangle|n\rangle|s\rangle \in {\cal H}_x\otimes {\cal H}_y\otimes {\cal H}_{\sf int}$. The pair of integers \((j,n)\) labels the magnetic unit cells and $s$ 
labels the sites in any one unit cell.  

Now we consider a lattice with the number of lattice sites in the $x\,(y)$ direction being $L_x=N Q\,(L_y)$ and set $t_x=t_y=-t$. To make the magnetic unit cell explicit, let us introduce the array 
\[
\hat{\Psi}_{j,n}^\dagger=\begin{bmatrix} c_{j_Q+1,n}^\dagger & c_{j_Q+2,n}^\dagger &\dots & c_{j_Q+Q,n}^\dagger\end{bmatrix}
\]
that collects creation operators of one unit cell. Then one can rewrite Eq.\,(\ref{Harper-Hofstadter}) as
\begin{align}\label{H_HH}
\widehat{H}_\delta=
-&t\sum_{j=1}^{N}\sum_{n=1}^{L_y}\left[\hat{\Psi}^\dagger_{j,n}
v\hat{\Psi}_{j,n}+\left(\hat{\Psi}_{j,n}^\dagger u\hat{\Psi}^{\;}_{j,n+1}+\text{H.c.}\right)\right]\nonumber\\
-&t\sum_{j=1}^{N-\delta}\sum_{n=1}^{L_y}\left(\hat{\Psi}_{j,n}^\dagger h_1\hat{\Psi}^{\;}_{j+1,n}+\text{H.c.}\right),\\
v=\hspace*{-0.1cm}\sum_{s=1}^{Q-1}&\left(\ket{s}\bra{s+1}+\text{H.c.}\right),\,u=\hspace*{-0.1cm}\sum_{s=1}^{Q} \omega^s|s\rangle\langle s|,\,h_1=|Q\rangle\langle 1|,\nonumber
\end{align}
where $v$, $u$ and $h_1$ are \(Q\times Q\) matrices associated with ${\cal H}_{\sf int}$, and $\delta\in\{0,1\}$ depending on the geometry.
\begin{itemize}
\item Cylinder geometry, \(\delta=1\). 
Given \(\hat{\Psi}_{j,L_y+1}=\hat{\Psi}_{j,1}\), \(\widehat{H}_1\) describes 
the Harper model rolled up into a cylinder coaxial with the \(x\) direction (see Fig.\,\ref{fig:Cyl}).
\item Torus geometry, \(\delta=0\). 
Given \(\hat{\Psi}_{j,L_y+1}=\hat{\Psi}_{j,1}\) and \(\hat{\Psi}_{N+1,n}=\hat{\Psi}_{1,n}\),
the Hamiltonian \(\widehat{H}_0\) describes the Harper model rolled up into a torus.
\end{itemize}

In terms of shift operators $V_{x,\delta}$ and $V_y$
\begin{align*}
	V_{x,\delta}=T_x+(1-\delta) |N\rangle \langle 1|,\quad V_y=T_y+|L_y\rangle \langle 1|,
\end{align*}
where $T_x=\sum_{j=1}^{N-1}|j\rangle\langle j+1|$ and $T_y=\sum_{n=1}^{L_y-1}|n\rangle\langle n+1|$, the single-particle Hamiltonian associated with $\widehat H_\delta$ is 
\begin{align*}
\hspace*{-0.1cm}H_\delta=-t\left[\mathds{1}\hspace*{-0.05cm}\otimes\hspace*{-0.05cm}\mathds{1}\hspace*{-0.05cm}\otimes\hspace*{-0.05cm}v +\big(\mathds{1}\hspace*{-0.05cm}\otimes\hspace*{-0.05cm}V_y\hspace*{-0.05cm}\otimes\hspace*{-0.05cm}u+V_{x,\delta}\hspace*{-0.05cm}\otimes\hspace*{-0.05cm}\mathds{1}\hspace*{-0.05cm}\otimes\hspace*{-0.05cm}h_1+\text{H.c.}\big)\right].
\end{align*}
For either cylinder or torus geometry, \(V_y\) and \(V^\dagger_y\) can be diagonalized simultaneously in the $k_y$ basis,
\begin{align*}
	|k_y\rangle=\frac{1}{\sqrt{L_y}}\sum_{n=1}^{L_y}e^{ik_yn}|n\rangle,\quad k_y=0,\frac{2\pi}{L_y},\dots,\frac{2\pi (L_y-1)}{L_y},
\end{align*}
and the 
single-particle Hamiltonian labeled by $k_y$ is
\begin{align}\label{hn}
\langle k_y|H_\delta |k_y\rangle=-t\left[\mathds{1}\otimes h_0+ \big(V_{x,\delta}\otimes h_1+ \text{H.c.}\big)\right],\\
h_0=v+(e^{ik_y}u+\text{H.c.})=v+\sum_{s=1}^Q\mu_s(k_y)|s\rangle\langle s|,\nonumber
\end{align}
where $\mu_s(k_y)=2\cos(k_y+2\pi\phi s)$. 
For torus geometry, \(V_{x,0}\) and $V_{x,0}^\dagger$ of Eq.\,(\ref{hn}) can be further diagonalized simultaneously in the $k_x$ basis 
\[
|k_x\rangle=\frac{1}{\sqrt{N}}\sum_{j=1}^{N}e^{iQk_xj}|j\rangle,\quad k_x=0,\frac{2\pi}{L_x},\dots,\frac{2\pi (N-1)}{L_x},
\]
and 
the single-particle Hamiltonian labeled by $k_x,\,k_y$ is
\begin{align}
\langle k_x,k_y|H_0 |k_x,k_y\rangle=-t\left[h_0+ (e^{iQk_x} h_1+ \text{H.c.}) \right],
\end{align}
which is just the Bloch Hamiltonian.

\subsection{Symmetries}\label{symmetries}

For \(Q>2\), the Harper model does not display any internal 
symmetries and it belongs to class A, which is topologically nontrivial in two dimensions. The case $Q=2$ is special since the matrix \(u\) is real-valued, rendering $H_\delta$ real, 
with time-reversal symmetry squared to one. 
It follows that, for \(Q=2\), the Harper model belongs to class AI, which is topologically trivial in two dimensions.

Now let us point out some accidental, non-internal, chiral symmetries of the Harper model that become manifest in the spectrum. The starting point is to notice that a shift operator like \(T_x\), \(T_y\), or \(T_{\sf int}=\sum_{s=1}^{Q-1}\ket{s}\bra{s+1}\) is \(U(1)\) covariant. Let us focus on \(T_y\) 
first. Rotations of \(T_y\) are generated by the position-like operator $R_y=\sum_{n=1}^{L_y}\, n \, |n\rangle\langle n|$ because \([R_y,T_y]=-T_y\). Similarly, $[R_y,T_y^\dagger]=T_y^\dagger$. Therefore, we have
\[
e^{i\theta R_y} T_y e^{-i\theta R_y}=e^{-i\theta}T_y,\quad 
e^{i\theta R_y} T_y^\dagger e^{-i\theta R_y}=e^{i\theta}T_y^\dagger.
\]
The periodic shift in the \(y\) direction can be decomposed as $V_y=T_y+(T_y^\dagger)^{L_y-1}$, which leads to $e^{i\theta R_y} V_y e^{-i\theta R_y}=e^{-i\theta}T_y+e^{i\theta(L_y-1)}(T_y^\dagger)^{L_y-1}$. Hence, for \(L_y\) even, we have
\[
e^{i\pi R_y} V_y e^{-i\pi R_y}=-V_y.
\]
Next let us 
focus on \(T_{\sf int}\). The position-like operator is $R_{\sf int}=\sum_{s=1}^Q\, s \, |s\rangle\langle s|$. From \(v=T^{\;}_{\sf int}+T_{\sf int}^\dagger\), we have
\[
e^{i\pi R_{\sf int}} v e^{-i\pi R_{\sf int}}=-v,
\]
and from \(h_1=(T_{\sf int}^\dagger)^{Q-1}\), we have for \(Q\) even
\[
e^{i\pi R_{\sf int}} h_1 e^{-i\pi R_{\sf int}}=-h_1.
\]
Putting all the pieces together, 
the unitary operator
\[
{\cal U}=e^{i\pi(\mathds{1}\otimes R_y\otimes \mathds{1}+ \mathds{1}\otimes \mathds{1}\otimes R_{\sf int})}
\]
anticommutes with the single-particle Hamiltonian $H_\delta$ for \(L_y\) and \(Q\) even, meaning that the spectrum of $H_\delta$ is symmetric with respect to zero. The even-odd effect in \(L_y\) can be avoided by making the boundary conditions open in the \(y\) direction. Then, the Harper model is chiral symmetric for all \(Q\) even. These results are independent of the boundary conditions in the \(x\) direction. Hence, the chiral symmetry will also be noticeable in the edge modes for \(Q\) even and larger than two. Because of this (non-internal) chiral symmetry in real space, it turns out that there exist superconductors that have the same Cantor set structure as the Harper model by duality transformations (see Appendix\,\ref{duality}). 
The duality must re-arrange the energy spectrum in momentum space in order to display the natural (internal) particle-hole symmetry of a superconductor.

\section{The Bloch Ansatz for the cylinder}
\label{BetheAnsatzSolutions}

In this section we will diagonalize, in closed form, the Harper model for 
the cylinder geometry (see Fig.\,\ref{fig:Cyl}) and rational flux per plaquette. As it turns out, it is advantageous to terminate the system
somewhere inside the last unit cell. At the same time, it is important
 to maintain the block structure of the single-particle Hamiltonian. 
The way forward is to add sites to the last unit cell that are completely decoupled from the Harper model.

To be precise, let the length of the cylinder be $L_x=QN-\nu$, where $\nu\in\{0,1,\dots,Q-1\}$ and $N$ is the number of magnetic 
unit cells in the $x$ direction.
If \(\nu=0\), then all the unit cells are complete. 
If $\nu>0$, we consider an extension of our system in the \(x\) direction to length $QN$ and subtract the hopping terms 
connecting the additional $\nu$ sites to neighboring sites.
We model this situation by splitting the single-particle Hamiltonian 
into two terms,  
\begin{align}
H=H_N+W,\nonumber
\end{align}
where $H_N$ is $\langle k_y|H_1|k_y\rangle$ of Eq.\,\eqref{hn}, describing the system with $QN$ sites in the $x$ direction, and $W$ models the special boundary conditions resulting from decoupling the last $\nu$ sites. For example, if $\nu=1$, then
\begin{align}\label{W}
W=|N\rangle\langle N|\otimes\left(t\ket{Q-1}\bra{Q}+\text{H.c.}\right).
\end{align}
Note that $H$ is now reducible and composed of $L_x\times L_x$ and $1\times 1$ matrix blocks.

\vspace*{-0.57cm}
\subsection{Diagonalization}
\label{Diagonalization}

To construct an Ansatz for states of the cylinder, we first split the eigenvalue equation $H \ket{\epsilon}=\epsilon \ket{\epsilon}$ into a system of two equations, the bulk  
and boundary equations 
\begin{align}
P_B H \ket{\epsilon} = \epsilon P_B\ket{\epsilon},\quad
P_\partial H \ket{\epsilon} = \epsilon P_\partial \ket{\epsilon},\nonumber
\end{align}
with $P_B=\sum_{j=2}^{N-1} \ket{j}\bra{j}\otimes \mathds{1}_Q$ the bulk projector and $P_\partial=\mathds{1}_{QN}-P_B$ the boundary projector. We then construct a basis of $2Q$ independent solutions of the bulk equation and use a linear combination of them as the Ansatz solution for the boundary equation. To achieve that, within the {\it generalized Bloch theorem} formulation \cite{PRL1,BlochAnsatz,PRB1,PRB2}, we consider the analytic continuation 
\begin{align}
    H(z)=-t\left(h_0+zh_1+z^{-1}h_1^\dagger\right)  \nonumber  
\end{align}
of the Bloch Hamiltonian $H(z=e^{iQk_x})$, and the associated characteristic polynomial 
\begin{align}
    P(z)= z^Q\det\left[H(z)-\epsilon \mathds{1}_Q\right].\nonumber
\end{align}
A nonzero vector
$|u(\epsilon,z)\rangle=[u_1(\epsilon,z) \cdots u_Q(\epsilon,z)]^\text{T}$ such that
\(
H(z)|u(\epsilon,z)\rangle =\epsilon|u(\epsilon,z)\rangle
\)
exists only if $P(z)=0$. Hence, we usually have for each \(\epsilon\) two distinct nonzero roots $z_\ell$, $\ell=1,2,$ with $z_1z_2=1$ but $z_\ell\neq \pm 1$. They are 
associated with two independent solutions of the bulk equation. In addition, there is 
a zero root with multiplicity $Q-1$ associated with the other $2(Q-1)$ independent solutions (see Ref.\,[\onlinecite{PRB1}] for details). Here we write down these independent solutions directly. They are
\begin{align}
\ket{\psi_\ell}&=\ket{z_\ell}\otimes|u(\epsilon,z_\ell)\rangle\quad\mbox{for}\ \ell=1,2,\nonumber\\
\ket{\psi_s^+}&=\ket{N}\otimes|s+1\rangle\quad\mbox{for}\ s=1,2,\dots,Q-1,\nonumber\\
\ket{\psi_s^-}&=\ket{1}\otimes|s\rangle\quad\mbox{for}\ s=1,2,\dots,Q-1,\nonumber
\end{align}
where $\ket{z_\ell}=\sum_{j=1}^N z_\ell^j \ket{j}$ and $|s\rangle$ is a $Q$-dimensional unit (column) vector with vanishing components except for the $s\text{th}$ component which is equal to one. A \textit{Bloch Ansatz} \(\ket{\epsilon}\) is a linear combination of these $2Q$ solutions of the bulk equation. As a formula,  
\begin{align}\label{Bethe-ansatz}
\ket{\epsilon}=
\sum_{\ell=1}^2\alpha_\ell\ket{\psi_\ell}+\sum_{s=1}^{Q-1}\left(\alpha_{2+s}\ket{\psi_s^+}+\alpha_{Q+1+s}\ket{\psi_s^-}\right).
\end{align}
The general \textit{Bloch Ansatz} of Eq.\,\eqref{Bethe-ansatz} can be plugged into the boundary equation to get a system of linear equations
$B(\epsilon)[\alpha_1\,\alpha_2\,\cdots\,\alpha_{2Q}]^\text{T}=0$,
with $B(\epsilon)$ a $2Q\times 2Q$ boundary matrix (see Ref.\,[\onlinecite{PRB1}]).
The case \(\nu=1\) is specially important. Then the boundary matrix is
\begin{widetext}
	\setcounter{MaxMatrixCols}{20}
	\begin{align}
	&B(\epsilon)=
	\begin{bmatrix}
	-u_Q(\epsilon,z_1)&-u_Q(\epsilon,z_2)					&0 	   &0	  &\dots &0  	 &0		   	 	   &f_1	  &1	 &0		&\dots	&0\\
	0		 &0		   					&0	   &0 	  &\dots &0 	 &0 		  	   &1	  &f_2	 &1		&\ddots	&\vdots\\
	0		 &0		   					&0	   &0 	  &\dots &0 	 &0 		 	   &0	  &1     &\ddots&\ddots &0\\
	\vdots	 &\vdots   					&\vdots&\vdots&\ddots&\vdots &\vdots     	   &\vdots&\ddots&\ddots&f_{Q-2}&1\\
	0		 &0		   					&0	   &0 	  &\dots &0 	 &0 			   &0	  &\dots &0		&1	   	&f_{Q-1}\\
	0		 &0		   					&0	   &0 	  &\dots &0 	 &0 		 	   &0	  &0	 &\dots	&0		&1\\
	0		 &0		   					&1	   &0	  &\dots &0	     &0 			   &0	  &0 	 &\dots &0 	 	&0\\
	0		 &0		   					&f_2   &1	  &0	 &\dots  &0 			   &0 	  &0 	 &\dots &0 		&0\\
	\vdots	 &\vdots   					&1	   &f_3   &\ddots&\ddots &\vdots	 	   &\vdots&\vdots&\ddots&\vdots &\vdots\\
	0		 &0		   					&0	   &\ddots&\ddots&1	     &0 		 	   &0	  &0 	 &\dots &0 	 	&0\\
	-z_1^Nu_Q(\epsilon,z_1)  &-z_2^Nu_Q(\epsilon,z_2)     &\vdots&\ddots&1	 &f_{Q-1}&0 		 	   &0	  &0     &\dots &0 	 	&0\\
	f_Qz_1^Nu_Q(\epsilon,z_1)&f_Qz_2^Nu_Q(\epsilon,z_2)   &0     &\dots &0     &0		 &f_Q  			   &0     &0     &\dots &0 	    &0
	\end{bmatrix},\nonumber
	\end{align}
in units of $-t$ and in terms of the shorthand notation 
\(
f_s=\epsilon/t+\mu_s(k_y)=\epsilon/t+2\cos(k_y+2\pi\phi s).
\) 

A Bloch Ansatz satisfies the boundary equation nontrivially for some \(\epsilon\) provided that $\text{det}B(\epsilon)=0$.  For our model in particular, and back
to the case with \(\nu\) arbitrary, 
\begin{align}
	\text{det}B(\epsilon)
	=(-1)^{Q-\nu+1}\bigg[\prod_{s=1}^\nu f_{Q-\nu+s}\bigg]\text{det}\begin{bmatrix}
	u_Q(\epsilon,z_1)				&u_Q(\epsilon,z_2)\\
	z_1^Nu_{Q-\nu+1}(\epsilon,z_1)	&z_2^Nu_{Q-\nu+1}(\epsilon,z_2)
	\end{bmatrix},\nonumber
\end{align}
\end{widetext}
with the understanding that, for the case $\nu=0$, $\prod_{s=1}^\nu f_{Q-\nu+s}=1$ and 
$z_\ell^Nu_{Q-\nu+1}(\epsilon,z_\ell)= z_\ell^{N+1}u_1(\epsilon,z_\ell)$, which is consistent with the case $L_x=Q(N+1)-\nu$ but with $\nu=Q$. 

In general, 
there are $\nu$ solutions associated with the vanishing of the product 
\[
\prod_{s=1}^\nu f_{Q-\nu+s}=0.
\] 
They correspond to states of  the last $\nu$ decoupled sites. The Ansatz solutions with $z_1z_2=1$ and
\begin{align}\label{det(B)arbitraryBC}
	\text{det}\begin{bmatrix}
	u_Q(\epsilon,z_1)				&u_Q(\epsilon,z_2)\\
	z_1^Nu_{Q-\nu+1}(\epsilon,z_1)	&z_2^Nu_{Q-\nu+1}(\epsilon,z_2)
	\end{bmatrix}=0
\end{align}
correspond to the other $QN-\nu$ sites.

When $\nu\neq 1$, Eq.\,(\ref{det(B)arbitraryBC}) does not admit a general decoupling between bulk and boundary states and therefore, for a finite size system, does not generally admit analytical solutions to either the bulk states or the boundary states. However, when $\nu=1$, Eq.\,(\ref{det(B)arbitraryBC}) 
becomes
\begin{align}\label{detcond}
(z_1^N-z_2^N)u_Q(\epsilon,z_1)u_Q(\epsilon,z_2)=0.
\end{align}
The solutions with $z_1^N=z_2^N$ correspond to bulk states, and the solutions with $u_Q(\epsilon,z_1)u_Q(\epsilon,z_2)=0$ usually (although not always) correspond to boundary states. In the following, we focus on the case $\nu=1$.

\subsubsection{Bulk states}

When $z_1^N=z_2^N$, we have $z_1=e^{i\pi j/N}$ and $z_2=e^{-i\pi j/N}$ ($j=1,2,\dots,N-1$), either of which could be plugged into the characteristic equation det$[H(z_\ell)-\epsilon\mathds{1}_Q]=0$, leading to the characteristic polynomial in $\epsilon$
\begin{align}\label{bulkpolynomial}
	&\text{det}[M^{(k_y)}_{0,Q}]-\text{det}[M^{(k_y)}_{1,Q-2}]-(-1)^Q(z_\ell+z_\ell^{-1})=0,\\
	&\text{or\quad}\prod_{k=1}^{Q}F^{-1}_k-\prod_{k=1}^{Q-2}G^{-1}_{k+1}-(-1)^Q(z_\ell+z_\ell^{-1})=0,\nonumber
\end{align}
where $M^{(k_y)}_{a,b}$ represent the $b\times b$ matrices
\begin{align}
	&M^{(k_y)}_{a,b}=\begin{bmatrix}
	f_{a+1}		&1		&0		&\dots		&0\\
	1		&f_{a+2}	&1		&\ddots		&\vdots\\
	0		&1		&\ddots	&\ddots		&0\\
	\vdots	&\ddots	&\ddots	&f_{a+b-1}	&1\\
	0		&\dots	&0		&1	   		&f_{a+b}
	\end{bmatrix},\nonumber
\end{align}
with the understanding that $\text{det}[M^{(k_y)}_{1,0}]=1$ in Eq.\,(\ref{bulkpolynomial}). In addition, $F_k$ is the $k_y$-dependent continued fraction
\begin{align}
F_k(k_y)=\cfrac{1}{f_k-\cfrac{1}{f_{k-1}-\cfrac{1}{\quad\ddots \raisebox{-0.9em}{\ensuremath{f_2-\cfrac{1}{f_1} }}}}},\nonumber
\end{align}
with $F_1(k_y)=1/f_1$, and $G_{k+1}(k_y)=F_k(k_y+2\pi\phi)$.
 
From the characteristic polynomial Eq.\,\eqref{bulkpolynomial} one determines 
the bulk eigenvalues $\epsilon\equiv \epsilon_{s,j}$. Notice 
that $s=1,2,\dots,Q$ represents the band index and $j=1,2,\dots,N-1$ labels the phase factors \(z_1=e^{i\pi j/N}=z_2^{-1}\). 
The corresponding bulk eigenstates are
\begin{align}
|\epsilon_{s,j}\rangle=|z_1\rangle\otimes|u(\epsilon_{s,j},z_1)\rangle-|z_2\rangle\otimes|u(\epsilon_{s,j},z_2)\rangle,\label{bulkstates}
\end{align}
which corresponds to the Ansatz solution of Eq.\,(\ref{Bethe-ansatz}) with only $\alpha_1=1$ and $\alpha_2=-1$ different 
from zero. The formula for the components of $|u(\epsilon,z_\ell)\rangle$ is
\begin{align}
	u_s(\epsilon,z_\ell)=\bigg[&(-1)^sz_\ell^{-1} \big(1+\sum_{\ell=1}^{Q-s-1}\prod_{k=\ell}^{Q-s-1}F_{Q-k}F_{Q-1-k}\big)\nonumber\\
	&\ \ \times \prod_{k=1}^{s}F_k +(-1)^{Q-s}\prod_{k=1}^{Q-s}F_{Q-k}\bigg]u_Q,\nonumber
\end{align}
for $s=1,2,\dots,Q-1$, with the understanding that
\[
u_{Q-1}(\epsilon,z_\ell)=\big[(-1)^{Q-1}z_\ell^{-1}\prod_{k=1}^{Q-1}F_k-F_{Q-1}\big]u_Q.
\]  
One can choose $u_Q$ to be some constant such that $|\epsilon_{s,j}\rangle$ is normalized.
If $u_Q$ is chosen to be purely (real) imaginary, then $|\epsilon_{s,j}\rangle$ is purely (imaginary) real.

\subsubsection{Boundary states}\label{boundarysolutions}

Back to Eq.\,\eqref{detcond} with $u_Q(\epsilon,z_1)u_Q(\epsilon,z_2)=0$. Since there is no difference between $u_Q(\epsilon,z_1)=0$ and $u_Q(\epsilon,z_2)=0$, either of them could be used to determine the other $Q-1$ states, which usually correspond to boundary states except when $Q=2$. Let us just focus on $u_Q(\epsilon,z_1)=0$. 
To have a non-vanishing eigenvector $|u(\epsilon,z_1)\rangle=[u_1(\epsilon,z_1) \cdots u_Q(\epsilon,z_1)]^\text{T}$ 
of $H(z_1)$, we must have $z_1 u_1(\epsilon,z_1)+u_{Q-1}(\epsilon,z_1)=0$ and 
\begin{align}\label{edgepolynomial}
\text{det}[M^{(k_y)}_{0,Q-1}]=0,
\end{align}
which is $\prod_{k=1}^{Q-1}F^{-1}_k=0$ and leads to the other eigenvalues $\epsilon\equiv \epsilon(r)$ ($r=1,2,\dots,Q-1$), together with
	\begin{align}
	u_s(\epsilon,z_1)=(-1)^{s-1}\bigg[\prod_{k=1}^{s-1}F^{-1}_k\bigg]u_1,\nonumber
	\end{align}
where $s=2,3,\dots,Q-1$, and $u_1$ is chosen to be some constant so that the associated eigenstate $|\epsilon(r)\rangle$ is normalized. 
Back to $z_1 u_1+u_{Q-1}=0$, we have
	\begin{align}\label{z1}
	z_1
	=(-1)^{Q-1}\prod_{k=1}^{Q-2}F^{-1}_k=(-1)^{Q-1}\text{det}[M^{(k_y)}_{0,Q-2}],
	\end{align} 
and finally we have the associated eigenstate as
	\begin{align}
	|\epsilon(r)\rangle=|z_1\rangle\otimes|u[\epsilon(r),z_1]\rangle,\label{edgestates}
	\end{align}
which corresponds to the Ansatz solution of Eq.\,(\ref{Bethe-ansatz}) with the non-vanishing $\alpha_\ell$ being $\alpha_1=1$. 

As we specified at the beginning of Sec.\,\ref{Diagonalization}, we have assumed that $z_1\neq \pm 1$. However, if it happens that $z_1=\pm 1$ at some special momenta $k_y$ where $\epsilon(r)$ touches the bulk band edges, 
then Eq.\,(\ref{edgestates}) will correspond to bulk states rather than boundary states. See Appendix\,\ref{degenerateroots} for quantitative discussions of the case $z_1=z_2=\pm 1$.

From Eq.\,(\ref{edgestates}), together with $\ket{z_1}=\sum_{j=1}^N z_1^j \ket{j}$ and $z_1\in\mathbb{R}$, we see immediately that $|\epsilon(r)\rangle$ will be localized at $x=1$ if $|z_1|<1$, and it will be localized at $x=QN-1$ if $|z_1|>1$. Furthermore, since $|u[\epsilon(r),z_1]\rangle=[u_1\quad \cdots\quad -z_1u_1\quad 0]^\text{T}$, we know that the probability of finding a boundary state at $x=Qj\,(j=1,2,\dots,N-1)$ vanishes, and the probability of finding a boundary state at either $x=Qj-1$ or $x=Qj+1$ is the same.

\subsection{A consequence of gauge invariance}
The bulk characteristic polynomial of Eq.\,(\ref{bulkpolynomial}) 
displays an interesting property: its roots (bulk eigenvalues) depend solely on the parameter $\lambda$ defined as
\begin{align}\label{lambda}
\lambda=\frac{z_\ell+z_\ell^{-1}}{2}+\cos(Qk_y).
\end{align} 
This can be shown as follows. Suppose we impose periodic boundary 
conditions in both the $x$ and $y$ directions. Then  $z_\ell=e^{iQk_x}$ and 
therefore Eq.\,(\ref{bulkpolynomial}) becomes
\begin{align}\label{ky}
\text{det}[M^{(k_y)}_{0,Q}]-\text{det}[M^{(k_y)}_{1,Q-2}]-(-1)^Q2\cos(Qk_x)=0.
\end{align}
Had we used a different gauge, the resulting eigenvalues should be identical.  
For example, had we used the Landau gauge $\mathbf{A}=-\frac{h}{e}\phi y\mathbf{e}_x$, and 
impose periodic boundary conditions in both directions, then 
Eq.\,(\ref{bulkpolynomial}) would become 
\begin{align}\label{kx}
\text{det}[M^{(k_x)}_{0,Q}]-\text{det}[M^{(k_x)}_{1,Q-2}]-(-1)^Q2\cos(Qk_y)=0,
\end{align}
effectively exchanging $x \leftrightarrow y$. 
Since the roots (and its multiplicities) of the characteristic polynomials and the coefficient 
of their leading $\epsilon^Q$ term 
are identical, 
Eqs.\,(\ref{ky}) and (\ref{kx}) represent the same polynomial.
Hence, 
the $k$-dependent term in the polynomial in $\epsilon$, $\text{det}[M^{(k)}_{0,Q}]-\text{det}[M^{(k)}_{1,Q-2}]$, must be equal to $-(-1)^Q2\cos(Qk)$.

Let us go back to the cylinder boundary conditions with terminations in the \(x\) direction and the bulk equation Eq.\,(\ref{bulkpolynomial}). Note that $\text{det}[M^{(k_y)}_{0,Q}]-\text{det}[M^{(k_y)}_{1,Q-2}]$ does not depend on $z_\ell$, and in view of what we have just shown regarding its $k_y$ dependence, it turns out that the characteristic polynomial Eq.\,(\ref{bulkpolynomial}) can only depend on the single parameter $-(-1)^Q[2\cos(Qk_y)+z_\ell+z_\ell^{-1}]=-(-1)^Q2\lambda$. When $\nu=1$, we have
\[
\lambda=\cos(\frac{\pi j}{N})+\cos(Qk_y)\equiv \lambda(j,k_y)\]
for the bulk states. We will use this notation in Sec.\,\ref{Examples}.

Similarly, for periodic boundary conditions in the $x$ and $y$ directions, we have the Bloch Hamiltonian $H(z_\ell=e^{iQk_x})$. Then, energy eigenvalues only depend on the parameter\cite{Thouless,Santos} $\lambda=\cos(Qk_x)+\cos(Qk_y)\equiv \lambda(k_x,k_y)$. More generally, for twisted boundary conditions in the $x$ direction and the periodic boundary condition in the $y$ direction (see Appendix\,\ref{TwistedBCs} for details), we have $z_\ell=e^{i(2\pi j+\Theta)/N}$ and $\lambda=\cos(\frac{2\pi j+\Theta}{N})+\cos(Qk_y)\equiv \lambda(j,\Theta,k_y)$, with $\lambda(j,0,k_y)=\lambda(k_x,k_y)$. Therefore, we see that bulk eigenvalues of the cylinder $\epsilon_{s,j}$ with $j$ (odd) even belong to exact eigenvalues of the torus with (anti-) periodic boundary conditions. Furthermore, from an eigenvalue problem standpoint, when $\lambda(j,\Theta,k_y)=\lambda(j,k_y)$, there is no difference between cylinder and torus geometries. 

\subsection{A Cantor set of the bulk spectrum}

The single-particle Harper Hamiltonian on the infinite cylinder  
[Hilbert space \(\ell^2(\mathds{Z}\times \{1,2,\dots,L_y\})\)] can be 
isometrically mapped by the Fourier transform to a block diagonal
operator \(\bigoplus_{k_y}H_{k_y}\). The blocks
\(H_{k_y}\) act naturally on the Hilbert space \(\ell^2(\mathds{Z})\). The eigenvalue equation for any of the \(H_{k_y}\) is a 
non-autonomous difference equation, sometimes called the quasi-Mathieu equation, having the flux $\phi$ and the 
ratio \(t_x/t_y\) as parameters. It was conjectured for a long time that the energy spectrum of \(H_{k_y}\) is a Cantor set for an irrational value of the flux\cite{Hofstadter}, and the conjecture 
was finally proven correct in Ref.\,[\onlinecite{Math}]. We would like to understand what are fingerprints 
of the Cantor set in the bulk spectrum of a finite cylinder. 

As a stepping stone, let us consider first 
the energy spectrum of the Harper model for the semi-infinite cylinder [Hilbert space 
\(\ell^2(\mathds{Z}^+\times \{1,2,\dots,L_y\})\)] and the rational flux
from the point of view of the generalized Bloch's theorem. The 
details are included in Appendix\,\ref{semi-infinite}.  
The conclusion is that the bulk energy spectrum of the semi-infinite cylinder (one termination) 
coincides with that of the infinite cylinder (no terminations) up to a finite number of spectral points: 
two bulk band energies are missing from the bulk spectrum of the semi-infinite cylinder. One can then 
build on the theorems about continuity of the spectrum with respect to the flux \cite{Math}, 
and conclude that the \textit{bulk} spectrum  of the \textit{semi-infinite} cylinder is also a Cantor 
set when the flux $\phi$ is an irrational number.

Let us consider next the finite cylinder (two terminations). 
We pick a sequence of fluxes $\phi_{\sf i}=P_{\sf i}/Q_{\sf i}\,({\sf i}\in\mathds{Z}^+)$ that 
approaches an irrational number $\phi_\infty$. Then, it can be shown\cite{Math} 
that, when ${\sf i}\rightarrow \infty$, the spectrum of the Harper model in the torus and in the thermodynamic limit, 
associated with $\phi_{\sf i}$ and $\lambda(j,\Theta,k_y)$, will approach a Cantor set 
which does not depend on $j$, $\Theta$ and $k_y$. Since $\lambda(j,k_y)\subset\lambda(j,0,k_y)\cup\lambda(j,\pi,k_y)$, 
the bulk spectrum of the Harper model in the finite cylinder and in the thermodynamic limit, associated with $\phi_{\sf i}$ and $\lambda(j,k_y)$, will 
also approach the Cantor set when ${\sf i}\rightarrow \infty$. 

Intuitively, the bulk spectrum of the finite cylinder approaches that of the semi-infinite one, and the bulk spectrum of the semi-infinite cylinder coincides with that of the infinite one up to a discrete number of points. Hence, one should be able to extract information about the Cantor set of the infinite cylinder from the finite cylinder after separating bulk and boundary spectra.

\subsection{An exact boundary theory}

In this subsection, we first present an exact boundary theory of the Harper model for $\nu=1$. Then, we generalize to the other cases when $\nu\neq 1$ in the limit $N\rightarrow \infty$.

\subsubsection{The case $\nu=1$}

As shown above, the case $\nu=1$ is special. 
For fixed $k_y$, there is an emergent $\mathfrak{u}(Q-1)$ symmetry, with $(Q-1)^2$ generators $\{d_r^\dagger d^{\;}_s|r,s=1,2,\dots,Q-1\}$. 
If one defines $c^\dagger_{m,k_y}=\frac{1}{\sqrt{L_y}}\sum_{n=1}^{L_y} e^{i k_y n}c^\dagger_{m,n}$ and rewrite Eq.\,(\ref{Harper-Hofstadter}) 
in terms of $c^\dagger_{m,k_y}$ as $\widehat{H}=\sum_{k_y} \widehat{H}_{1}(k_y)$ with 
\begin{align}\label{CylinderHamiltonian}
	\widehat{H}_{1}(k_y)=&-t\sum_{m=1}^{QN-1}\mu_m(k_y) c_{m,k_y}^\dagger c^{\;}_{m,k_y}\nonumber\\
	&-t\sum_{m=1}^{QN-2}\big(c_{m,k_y}^\dagger c^{\;}_{m+1,k_y}+\text{H.c.}\big)
\end{align}
and $\mu_m(k_y)=2\cos(k_y+2\pi\phi m)$, then the modes $d_r^\dagger$ satisfy $[\widehat{H}_{1}(k_y),d_r^\dagger]=\epsilon (r)d_r^\dagger$. Here,  $\epsilon(r)$ 
has no $N$ dependence [they are eigenvalues of Eq.\,(\ref{edgepolynomial})] and $d_r^\dagger$ are written as linear 
combinations of the $k_y$-dependent fermionic or bosonic creation operators as
\begin{eqnarray}
	d_r^\dagger=\sum_{j=1}^{N}\sum_{s=1}^{Q-1}z_1^{j+1}u_s[\epsilon(r),z_1]c_{j_Q+s,k_y}^\dagger,\nonumber
\end{eqnarray}
with $z_1$ and $u_s[\epsilon(r),z_1]$ associated to the $z_1$ and $|u[\epsilon(r),z_1]\rangle$ of Eq.\,(\ref{edgestates}). 
Because of this emergent $\mathfrak{u}(Q-1)$ symmetry, 
there is usually one boundary state in each of the $Q-1$ bulk 
band gaps, together with $N-1$ bulk states in each of the $Q$ bulk bands, which leads to a total of $(Q-1)+Q(N-1)=QN-1$ states as expected 
(see Fig.\,\ref{Fig_p/q}\,(c) and (f) for examples). 

Most importantly, since the energy spectrum of boundary states $\epsilon(r)$ no longer depends on the system size $N$, we are able to construct an exact boundary theory describing the boundary physics of the finite size Harper model. 
Based on the boundary characteristic polynomial, Eq.\,(\ref{edgepolynomial}), we see that boundary energies are eigenvalues of the following Hermitian matrix
\begin{align}
	&H_\partial(k_y)=
	-t\begin{bmatrix}
		\mu_1(k_y)	&1		    &0		&\dots		    &0\\
		1		    &\mu_2(k_y)	&1		&\ddots		    &\vdots\\
		0		    &1		    &\ddots	&\ddots		    &0\\
		\vdots	    &\ddots	    &\ddots	&\mu_{Q-2}(k_y)	&1\\
		0		    &\dots	    &0		&1	   		    &\mu_{Q-1}(k_y)
	\end{bmatrix} .
	\nonumber
\end{align}
In this sense, this matrix can be seen as the exact single-particle boundary Hamiltonian. In second quantization,
\begin{align}\label{BoundaryHamiltonian}
	\widehat{H}_\partial(k_y)=&-t\sum_{s=1}^{Q-1}\mu_s(k_y) c_{s,k_y}^\dagger c^{\;}_{s,k_y}\nonumber\\
	&-t\sum_{s=1}^{Q-2}\big(c_{s,k_y}^\dagger c^{\;}_{s+1,k_y}+\text{H.c.}\big).
\end{align}
Compared with Eq.\,(\ref{CylinderHamiltonian}), one immediately appreciates that the boundary Hamiltonian Eq.\,(\ref{BoundaryHamiltonian}) corresponds to the cylinder geometry of Fig.\,\ref{fig:Cyl} with $Q-1$ sites along the $x$ direction\cite{HatsugaiPRB}, describing $Q-1$ boundary states localized at either $x=1$ or $x=L_x$, and can be simply seen as a one-dimensional $(Q-1)$-band system labeled by $k_y$. 
 
Given that the energy spectrum of boundary states is neatly described by a quadratic fermionic Hamiltonian in one less dimension, it is natural to ask whether it is topologically trivial. Note that $H_\partial(k_y)$ is a real-valued matrix and the associated $s$th band single-particle eigenstates $|\psi_s(k_y)\rangle$ 
can be chosen as  real-valued as well. Thus, the Berry connection $\mathcal{A}_s(k_y)=i\langle\psi_s(k_y)|\partial_{k_y}|\psi_s(k_y)\rangle$ 
is purely imaginary. Since the Berry connection $\mathcal{A}_s(k_y)$ must be  real-valued, we therefore have $\mathcal{A}_s(k_y)=0$ 
and the Berry phase $\gamma_s=\int_{-\pi}^{\pi} dk_y\mathcal{A}_s(k_y)=0$, which implies that the one-dimensional boundary theory is topologically trivial. 
This is consistent with the tenfold way classification of free fermions \cite{Ryu} or the threefold way classification of free bosons \cite{Qiaoru2020}, 
where a one-dimensional system in the real symmetry class AI (with time-reversal symmetry squared to one) is topologically trivial. Note that the origin of the topological triviality of this boundary theory roots in the fact that it describes simultaneously both two edges (rather than one edge), which comprise time-reversal pairs of counter-propagating boundary states localized at different edges and with opposite momenta. However, the original boundary state localized at either edge is topologically nontrivial in general.

Does the boundary ($\phi$-dependent) energy spectra resemble that of Hofstadter's butterfly? The answer would be yes if the boundary energy spectrum were the exact complement of the bulk one. However, for an arbitrary $\phi$, this is not the case. For example, for a rational $\phi$, the bulk bands (plotted as a function of $k_y$) always overlap with the boundary energy spectrum and, thus, the latter is more than the 
complement of the bulk spectrum [see Fig.\,\ref{Fig_p/q}\,(c) for example]. In addition, boundary states can also exist in the region where there is no bulk band gap  [see $\epsilon(2)$ of Fig.\,\ref{Fig_p/q}\,(f) for example].
	
Similarly, one can ask whether the boundary energy spectrum in an irrational flux limit displays signatures of a Cantor set. Obviously, it is not a Cantor set itself because it has a nonzero measure. However, is it the complement of a Cantor set? The answer is probably yes, with the understanding that we have already excluded from the boundary energy spectrum those exceptional bulk energy points associated with $z_1=\pm 1$ in Eq.\,(\ref{edgestates}). That is because in the irrational flux limit, the bulk bands become flat so that the boundary energy spectrum will not overlap them. Note that the complement of a Cantor set is gapless, leading to the conjecture that the boundary energy spectrum in the limit $Q\rightarrow \infty$ is gapless.

\subsubsection{The case $\nu\neq 1$}

As mentioned before, when $\nu\neq 1$, the energy spectrum of boundary states depends on the system size $N$. In the limit $N\rightarrow \infty$, as shown in Appendix\,\ref{nuneq1}, 
the total number of boundary states is conserved. Those localized at $x=1$ (associated with $|z_1|<1$) can be determined from the boundary Hamiltonian $H_\partial(k_y)$, while those localized at $x=L_x$ (associated with $|z_1|>1$) are obtained from $H_\partial\big(k_y+2\pi\phi(1-\nu)\big)$. It just happens that, when $\nu=1$, $H_\partial(k_y)$ describes all boundary states simultaneously.

Physically, this can be understood as follows. In the limit $N\rightarrow \infty$, boundary conditions at $x=L_x$ usually do not influence boundary states localized at $x=1$, which guarantees that $H_\partial(k_y)$ describes at least half of all boundary states. However, boundary conditions at $x=L_x$ do influence boundary states localized at $x=L_x$. To understand the difference, let us look at Eq.\,(\ref{CylinderHamiltonian}) and its generalization
\begin{align}\label{CylinderHamiltonian_nu}
\widehat{H}_{\nu}(k_y)=&-t\sum_{m=1}^{QN-\nu}\mu_m(k_y) c_{m,k_y}^\dagger c^{\;}_{m,k_y}\nonumber\\
&-t\sum_{m=1}^{QN-\nu-1}\big(c_{m,k_y}^\dagger c^{\;}_{m+1,k_y}+\text{H.c.}\big).
\end{align}
Instead of labeling the $QN-\nu$ lattice sites by $m\in\{1,2,\dots,QN-\nu\}$ as shown above, if we relabel lattice sites by $m^\prime\in\{\nu,\nu+1,\dots,QN-1\}$, the physics should not change. After the relabeling, although boundary conditions at $x=1$ have changed, boundary conditions at $x=QN-1$ are the same as those of Eq.\,(\ref{CylinderHamiltonian}), and therefore, in the limit $N\rightarrow \infty$, boundary states localized at $x=QN-1$ should be the same as those of Eq.\,(\ref{CylinderHamiltonian}) as well. Combining the fact that $\mu_m(k_y)=\mu_{m+\nu-1}\big(k_y+2\pi\phi(1-\nu)\big)$ and $m+\nu-1\in\{\nu,\nu+1,\dots,QN-1\}$, we see immediately that, in the limit $N\rightarrow \infty$, boundary states localized at $x=QN-\nu$ associated with the original Eq.\,(\ref{CylinderHamiltonian_nu}) are exactly the same as those localized at $x=QN-1$ associated with Eq.\,(\ref{CylinderHamiltonian}) but shifted from $k_y$ to $k_y+2\pi\phi(\nu-1)$ because of relabeling of lattice sites.
	
\subsection{Some Examples}\label{Examples}

\subsubsection{The flux $\phi=1/2$}\label{p/q=1/2}
We have Eqs.\,(\ref{bulkpolynomial}) and (\ref{edgepolynomial}) as 
\begin{align}
\epsilon^2-2\lambda(j,k_y)-4=0,\nonumber\\
\epsilon-2\cos(k_y)=0,\nonumber
\end{align}
where $\epsilon$ is written in units of $t$. 
The closed-form single-particle energies are obtained as follows
\begin{align}
& \epsilon_{s,j}=(-1)^s\sqrt{4+2\lambda(j,k_y)},\label{eq.1/2-1}\\
& \epsilon(1)=2\cos k_y,\label{eq.1/2-2}
\end{align}
where $s=1,2$, $j=1,2,\dots,N-1$. Correspondingly, from Eq.\,(\ref{bulkstates}) we have (unnormalized) bulk states for Eq.\,(\ref{eq.1/2-1})
\begin{align}
|\epsilon_{s,j}\rangle=
|z_1\rangle\otimes
\begin{bmatrix}
-z_1^{-1}-1\\
f_1
\end{bmatrix}
-|z_2\rangle\otimes
\begin{bmatrix}
-z_2^{-1}-1\\
f_1
\end{bmatrix},\nonumber
\end{align}
where $z_1=e^{i\pi j/N}=z_2^{-1}$, $\ket{z_\ell}=\sum_{j=1}^N z_\ell^j \ket{j}$. From Eq.\,(\ref{edgestates}), the (unnormalized) eigenstate for Eq.\,(\ref{eq.1/2-2}) is
\begin{align}
|\epsilon(1)\rangle=|z_1=-1\rangle\otimes
\begin{bmatrix}
1\\
0
\end{bmatrix},\label{ES.1/2}
\end{align}
which is delocalized and merges with bulk states in the thermodynamic limit $N\rightarrow \infty$. The absence of localized states here is consistent with the analysis of Sec.\,\ref{symmetries} that the case $Q=2$ is topologically trivial.

\subsubsection{The flux $\phi=1/3$}\label{p/q=1/3}
We have Eqs.\,(\ref{bulkpolynomial}) and (\ref{edgepolynomial}) as 
\begin{align}
\epsilon^3-6\epsilon+2\lambda(j,k_y)=0,\nonumber\\
\epsilon^2-2\cos(k_y)\epsilon-4\sin^2 k_y=0,\nonumber
\end{align}
where $\epsilon$ is written in units of $t$. The closed-form single-particle energies are obtained as follows
\begin{align}
& \epsilon_{s,j}=\sqrt{8}\cos\frac{2\pi s+\arccos\frac{\lambda(j,k_y)}{-2\sqrt{2}}}{3},\label{eq.1/3-1}\\
& \epsilon(r)=\cos k_y+\cos(\pi r)\sqrt{1+3\sin^2 k_y},\label{eq.1/3-2}
\end{align}
where $s=1,2,3$, $j=1,2,\dots,N-1$, and $r=1,2$. Correspondingly, from Eq.\,(\ref{bulkstates}) we have (unnormalized) bulk states for Eq.\,(\ref{eq.1/3-1}) as
\begin{align}
|\epsilon_{s,j}\rangle=
|z_1\rangle\otimes
\begin{bmatrix}
-z_1^{-1}f_2+1\\
z_1^{-1}-f_1\\
f_2f_1-1
\end{bmatrix}
-|z_2\rangle\otimes
\begin{bmatrix}
-z_2^{-1}f_2+1\\
z_2^{-1}-f_1\\
f_2f_1-1
\end{bmatrix},\nonumber
\end{align}
where $z_1=e^{i\pi j/N}=z_2^{-1}$, $\ket{z_\ell}=\sum_{j=1}^N z_\ell^j \ket{j}$. From Eq.\,(\ref{edgestates}) we have (unnormalized) boundary states for Eq.\,(\ref{eq.1/3-2}) as
\begin{align}
|\epsilon(r)\rangle=|z_1\rangle\otimes
\begin{bmatrix}
1\\
-z_1\\
0
\end{bmatrix},\label{ES.1/3}
\end{align}
where
\begin{align}
z_1=f_1=-\sqrt{3}\sin k_y+\cos(\pi r)\sqrt{1+3\sin^2 k_y}.\label{z1.1/3}
\end{align}

\begin{figure*}[t]
	\includegraphics[width=\textwidth]{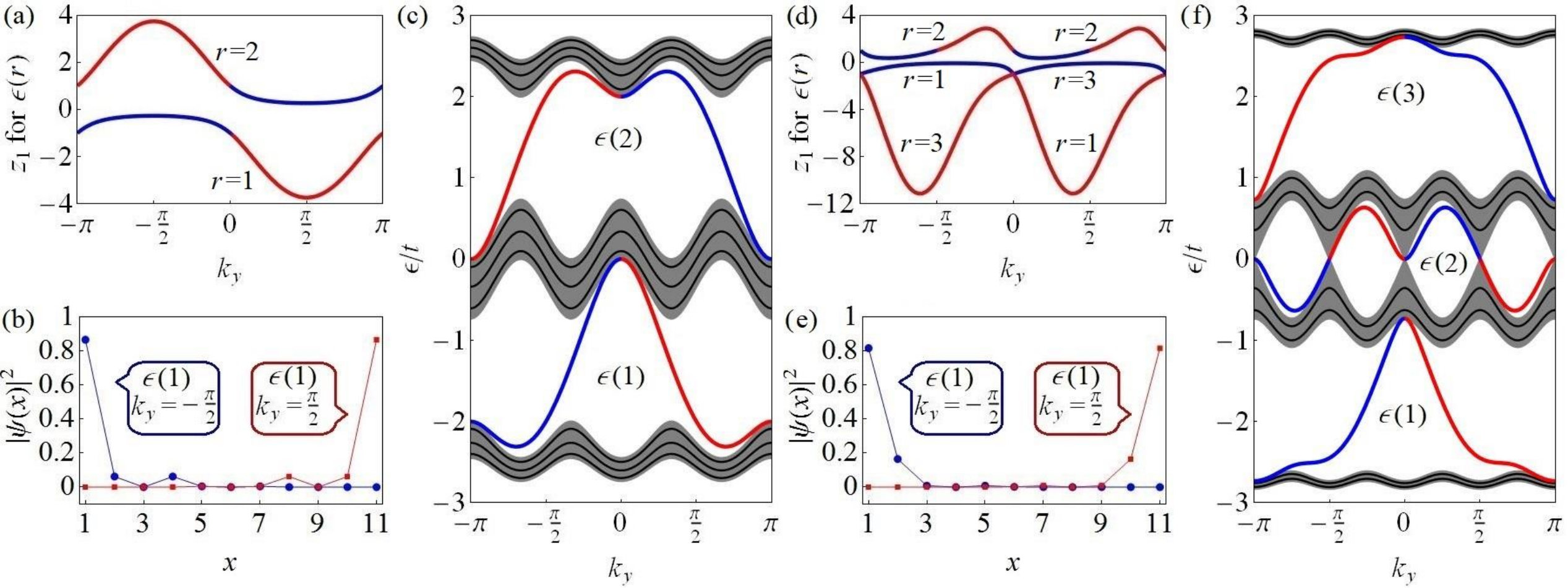}
	\caption{Analytical results of the Harper model in the cylinder with $L_x=3\times4-1=11$ for (a)-(c) $\phi=1/3$ and (d)-(f) $\phi=1/4$. The profile of $z_1$ associated with boundary states (a) of Eq.\,(\ref{ES.1/3}) and (d) of Eq.\,(\ref{ES.1/4}), where the blue lines are associated with $0<|z_1|<1$ and the red lines are associated with $|z_1|>1$. The probability distribution of two representative boundary states associated with $\epsilon(1)$ at $k_y=\{-\pi/2,\pi/2\}$ (b) of Eqs.\,(\ref{eq.1/3-2})-(\ref{ES.1/3}) and (e) of Eqs.\,(\ref{eq.1/4-2})-(\ref{ES.1/4}). The single-particle energy spectra associated with (c) Eqs.\,(\ref{eq.1/3-1})-(\ref{eq.1/3-2}) and (f) Eqs.\,(\ref{eq.1/4-1})-(\ref{eq.1/4-2}). The blue lines are associated with boundary states localized at $x=1$, while the red lines are associated with boundary states localized at $x=L_x$. Bulk spectra are indicated by black lines and, when $L_x=QN-1\rightarrow \infty$, they extend to form bulk bands colored in gray, although boundary spectra do not change.}
	\label{Fig_p/q}
\end{figure*}

To visualize these analytical results, we plot Eq.\,(\ref{z1.1/3}) in Fig.\,\ref{Fig_p/q}\,(a), where the blue colored segments are associated with $0<|z_1|<1$ and the red colored ones are associated with $|z_1|>1$. As we mentioned in Sec.\,\ref{boundarysolutions}, the blue colored segments are associated with Eq.\,(\ref{ES.1/3}) localized at $x=1$ and the red colored ones are associated with Eq.\,(\ref{ES.1/3}) localized at $x=L_x$. Note that there are two special momenta, $k_y=\{-\pi,0\}$, at which we have $z_1=(-1)^r$ and Eq.\,(\ref{ES.1/3}) corresponds to bulk states. In Fig.\,\ref{Fig_p/q}\,(b), we plot the probability distribution of two representative boundary states associated with $\epsilon(1)$, one at $k_y=-\pi/2$ where $z_1=-(2-\sqrt{3})$, while the other one at $k_y=\pi/2$ where $z_1=-(2+\sqrt{3})$. Although their energies of Eq.\,(\ref{eq.1/3-2}) have no system size $N$ dependence, the boundary states themselves are not isolated from the bulk and have tails in the bulk. Finally, we plot Eqs.\,(\ref{eq.1/3-1}) and (\ref{eq.1/3-2}) in Fig.\,\ref{Fig_p/q}\,(c), where the black lines are associated with bulk states and extend to form bulk bands colored in gray when $N\rightarrow \infty$, while the blue and red lines are associated with boundary states localized at $x=1$ and $x=L_x$ respectively. As we can see, there are two boundary states in the bulk band gaps. Using the argument of Laughlin and Halperin\cite{Laughlin,Halperin}, the Hall conductance $\sigma_{xy}=(-)e^2/h$ when the Fermi energy lies in one of the bulk band gaps.

\subsubsection{The flux $\phi=1/4$}\label{p/q=1/4}
We have Eqs.\,(\ref{bulkpolynomial}) and (\ref{edgepolynomial}) as 
\begin{align}
\epsilon^4-8\epsilon^2-2\lambda(j,k_y)+4=0,\nonumber\\
\epsilon^3-2\cos(k_y)\epsilon^2-(2+4\sin^2 k_y)\epsilon+8\sin^2 k_y\cos k_y=0,\nonumber
\end{align}
where $\epsilon$ is written in units of $t$ as well. The closed-form single-particle energies are obtained as follows
\begin{widetext}
	\begin{align}
	&\epsilon_{s,j}=2\cos\frac{\pi(2s+1)}{4}\sqrt{2+\cos\frac{\pi(2s-1)}{4}\sqrt{\lambda(j,k_y)+6}},\label{eq.1/4-1}\\
	& \epsilon(r)=\frac{2}{3}\cos k_y+\sqrt{8-\frac{32}{9}\cos^2 k_y}\,\cos\frac{2\pi r+\arccos\frac{40 \cos^3 k_y-27 \cos k_y}{\sqrt{2 (9-4\cos^2 k_y)^3}}}{3},\label{eq.1/4-2}
	\end{align}
where $s=1,2,3,4$, $j=1,2,\dots,N-1$, and $r=1,2,3$. Correspondingly, from Eq.\,(\ref{bulkstates}) we have (unnormalized) bulk states for Eq.\,(\ref{eq.1/4-1}) as
\begin{align}
|\epsilon_{s,j}\rangle=
|z_1\rangle\otimes
\begin{bmatrix}
-z_1^{-1}(f_3f_2-1)-1\\
z_1^{-1}f_3+f_1\\
-z_1^{-1}-(f_2f_1-1)\\
f_3f_2f_1-f_3-f_1
\end{bmatrix}
-|z_2\rangle\otimes
\begin{bmatrix}
-z_2^{-1}(f_3f_2-1)-1\\
z_2^{-1}f_3+f_1\\
-z_2^{-1}-(f_2f_1-1)\\
f_3f_2f_1-f_3-f_1
\end{bmatrix},\nonumber
\end{align}
\end{widetext}
where $z_1=e^{i\pi j/N}=z_2^{-1}$, $\ket{z_\ell}=\sum_{j=1}^N z_\ell^j \ket{j}$. From Eq.\,(\ref{edgestates}) we have (unnormalized) boundary states for Eq.\,(\ref{eq.1/4-2}) as
\begin{align}
|\epsilon(r)\rangle=|z_1\rangle\otimes
\begin{bmatrix}
1\\
-f_1\\
-z_1\\
0
\end{bmatrix},\label{ES.1/4}
\end{align}
where 
\begin{align}
z_1=-(f_2f_1-1)=-f_1/f_3=-\frac{\epsilon(r)-2\sin k_y}{\epsilon(r)+2\sin k_y}.\label{z1.1/4}    
\end{align}

Similar as the case of $\phi=1/3$, we plot Eq.\,(\ref{z1.1/4}) in Fig.\,\ref{Fig_p/q}\,(d), where the blue colored segments are associated with Eq.\,(\ref{ES.1/4}) localized at $x=1$ with $0<|z_1|<1$ and the red colored ones are associated with Eq.\,(\ref{ES.1/4}) localized at $x=L_x$ with $|z_1|>1$. Note that there are two special values of $k_y=\{-\pi, 0\}$ at which $z_1=-1$ for all $\epsilon(r)$, and another two special values of $k_y=\{-\pi/2, \pi/2\}$ at which $z_1=1$ for $\epsilon(2)$. At these values, Eq.\,(\ref{ES.1/4}) corresponds to bulk states. We also plot the probability distribution of two representative boundary states in Fig.\,\ref{Fig_p/q}\,(e), which are associated with $\epsilon(1)$ at $k_y=-\pi/2$ where $z_1=-(5-2\sqrt{6})$ and at $k_y=\pi/2$ where $z_1=-(5+2\sqrt{6})$. Each boundary state is not isolated from the bulk and has a tail in the bulk, although its energy Eq.\,(\ref{eq.1/4-2}) has no $N$ dependence. Finally, Eqs.\,(\ref{eq.1/4-1}) and (\ref{eq.1/4-2}) are plotted in Fig.\,\ref{Fig_p/q}\,(f), where the black lines are associated with bulk states and extend to form bulk bands colored in gray when $N\rightarrow \infty$, while the blue and red lines are associated with boundary states localized at $x=1$ and $x=L_x$ respectively. 
As we can see, the two middle bulk bands touch with each other at four nonequivalent Dirac points $k_y=\{-\pi, -\pi/2, 0, -\pi/2\}$ with zero energy in the thermodynamic limit. Nonetheless, the Hall conductance $\sigma_{xy}=(-)e^2/h$ when the Fermi energy lies in the bottom or the top bulk band gap.

\subsubsection{The flux $\phi=P/Q$ with $Q\geq5$}\label{q>=5}
Although we have the bulk states Eq.\,(\ref{bulkstates}) decoupled with the boundary states Eq.\,(\ref{edgestates}), not all states would admit closed-form solutions for fluxes with $Q\geq 5$. For example, when $\phi=1/5$, we have Eq.\,(\ref{bulkpolynomial}) as
\begin{align}
\epsilon^5-10\epsilon^3+\frac{5(7-\sqrt{5})}{2}\epsilon+2\lambda(j,k_y)=0,\nonumber
\end{align}
where $\epsilon$ is written in units of $t$. When 
$\lambda(j,k_y)=1/2$, after rationalization, the above quintic equation becomes an irreducible tenth degree polynomial equation, 
with its Galois group being $S_5\wr C_2$ (the wreath product of $S_5$ by $C_2$), which is unsolvable. Thus, the above quintic 
equation is not solvable in general. However, for $Q=5$, closed-form solutions for boundary states are still available because 
Eq.\,(\ref{edgepolynomial}) now becomes a quartic equation. Furthermore, for $Q=6$ or $Q=8$, Eq.\,(\ref{bulkpolynomial}) is 
still solvable because terms with odd degrees all vanish and it becomes an effective cubic equation or quartic equation.

\section{A Bethe Ansatz for the cylinder and torus geometries}\label{Berus}

In this section, we 
introduce a non-commutative algebra that is a deformation of the well-known Weyl algebra, of relevance in formulations of quantum mechanics \cite{Schwinger} and clock models \cite{Ortiz2012}. After introducing this algebra, we show  that it yields, in a natural way, representations of the group of magnetic translations and solutions of the Yang-Baxter equation. Finally, we show that 
our {\it bulk} Hamiltonian with $\lambda=0$ of Eq.\,(\ref{lambda}) can be diagonalized in terms of a Bethe Ansatz. 

\subsection{A deformed Weyl algebra}

Our deformed Weyl algebra features two parameters: a complex number \(z\) and a unimodular complex number \(\omega=e^{i2\pi\phi}\). 
The generators satisfy (not all independent) relations 
\begin{eqnarray}
U U^\dagger&=&\mathds{1}, \quad V_+ V_- = V_- V_+=\mathds{1},\nonumber\\
U^Q&=&\mathds{1}, \quad V_+^Q=z, \quad V_-^Q=z^{-1}, \nonumber\\
V_+U&=&\omega UV_+, \quad V_+U^\dagger=\omega^* U^\dagger V_+, \nonumber \\
V_-U&=&\omega^* UV_-, \quad V_-U^\dagger=\omega U^\dagger V_-. 
\label{clock_edf}
\end{eqnarray}
When the deformation parameter $z^*=z^{-1}$, $V^\dagger_+=V_-$, and if, in addition, 
$z=1$ the algebra reduces to the usual Weyl algebra. Note that the connection of the deformed Weyl algebra to the Yang-Baxter equation is presented in Appendix\,\ref{YBE}.

A $Q$-dimensional representation of this algebra is given by the $Q\times Q$ complex matrices
\begin{align}
U&=\sum_{s=1}^{Q} \omega^{s-1}|s\rangle\langle s|,\nonumber\\
{V_+}(z)&=z\ket{Q}\bra{1}+\sum_{s=1}^{Q-1}\ket{s}\bra{s+1},\nonumber\\
{V_-}(z)&=z^{-1}\ket{1}\bra{Q}+\sum_{s=1}^{Q-1}\ket{s+1}\bra{s}.
\end{align}
In the orthonormal basis $\ket{s}$ where $U$ is diagonal, the $V_\pm(z)$ act as shift operators 
\begin{eqnarray}
V_+\ket{s+1}=\ket{s}\,(s=1,2,\dots,Q-1),\,\,&V_+\ket{1}=z\ket{Q},\nonumber\\
V_-\ket{s}=\ket{s+1}\,(s=1,2,\dots,Q-1),\,\,&V_-\ket{Q}=z^{-1}\ket{1}.\nonumber
\end{eqnarray}
The basis that diagonalizes the $V_\pm(z)$ is
\begin{eqnarray}\hspace*{-0.5cm}
\ket{r} = \frac{1}{\sqrt{Q}} \sum_{s=1}^{Q}\omega^{(r-1)\cdot(s-1)} z^{\frac{s}{Q}}\ket{s}\quad(r=1,2,\dots,Q),
\label{eigenvofV1}
\end{eqnarray}
with $V_+\ket{r}=\omega^{r-1}z^{\frac{1}{Q}}\ket{r}$ 
and $V_-\ket{r}=(\omega^*)^{r-1}z^{-\frac{1}{Q}}\ket{r}$.
It is an orthonormal basis for $|z|=1$. In this basis it is the $U, U^\dagger$ that 
act as ladder operators. Concretely,  
\begin{eqnarray}
U\ket{r}=\ket{r+1}\,(r=1,2,\dots,Q-1),\quad U\ket{Q}=\ket{1},\nonumber \\
U^\dagger\ket{r+1}=\ket{r}\,(r=1,2,\dots,Q-1),\quad U^\dagger \ket{1}=\ket{Q}.\nonumber
\end{eqnarray}

\subsection{A Bethe Ansatz at \(\lambda=0\)}

It is straightforward to show that our {\it bulk} Hamiltonian is an element 
of the deformed Weyl algebra because
\begin{eqnarray}
 H(z)= -t\left[\omega_{k_y} U + \omega^*_{k_y} U^\dagger + V_+(z)+ V_-(z)\right]
\end{eqnarray}
where $\omega_{k_y}=e^{i k_y} \omega$. This bulk Hamiltonian displays a chiral symmetry
\begin{eqnarray}
{\cal U}_c= U_{\;}^{\frac{Q}{2}} V_+(z)^{\frac{Q}{2}}, 
\end{eqnarray}
for $Q$ even and arbitrary value of $\lambda=\frac{z+z^{-1}}{2}+\cos(Qk_y)$, such that $\{H(z), {\cal U}_c\}=0$. In particular, the spectrum of $H(z)$ is symmetric about the zero value. Consequently, if there exists a bulk zero-energy mode, then its degeneracy is always an even number.  

This chiral symmetry breaks down for $Q$ odd. However, at the special value $\lambda=0$, i.e., $z= - e^{i Q k_y}$, a new chiral symmetry emerges. To discover this symmetry, it is convenient to show first that the eigenvalues of $H(z=-e^{i Q k_y})$ do not depend on $k_y$ even though the eigenvectors do depend on it. One can show that there exists a unitary map 
\begin{eqnarray}
 \tilde H = {\cal U}^\dagger_{k_y} H(-e^{i Q k_y}) \, {\cal U}^{\;}_{k_y} ,
\end{eqnarray}
such that 
\begin{eqnarray}\hspace*{-0.5cm}
 \tilde H = -t\left[U + U^\dagger - V_+((-1)^{Q+1})- V_-((-1)^{Q+1})\right],\label{Hamilnok}
\end{eqnarray}
independent of $k_y$-dependence. Hence, for $Q$ odd, $\tilde H$ is an element of the standard Weyl algebra \cite{Ortiz2012}. In that context one learns that the discrete Fourier transform
\begin{eqnarray}\hspace*{-0.5cm}
{F}^{\dagger}=\frac{1}{\sqrt{Q}}
\begin{pmatrix}
1& 1& 1& \cdots& 1\\
1& \omega& \omega^2& \cdots& \omega^{Q-1}\\
1& \omega^2& \omega^4& \cdots& \omega^{2(Q-1)}\\
\vdots& \vdots& \vdots&      &\vdots \\
1& \omega^{Q-1}& \omega^{(Q-1)2}& \cdots& \omega^{(Q-1)(Q-1)}
\end{pmatrix} ,
\label{DFTmatrix}
\end{eqnarray} 
maps 
\begin{equation}
{F} U {F}^\dagger=V_-(1),\qquad {F} V_+(1) {F}^\dagger=U.
\label{wga_aut}
\end{equation}
It follows that \(F\) represents the chiral symmetry for $\tilde H$, i.e., 
$\{H(-e^{i Q k_y}),{\cal U}_{c,  {\rm odd}}\}=0$, with ${\cal U}_{c, {\rm odd}}=
{\cal U}_{k_y} F^\dagger {\cal U}^\dagger_{k_y}$. Notice that since $Q$ is odd, when $\lambda=0$, the existence of this symmetry implies that there exists at least one zero (bulk) energy mode (whose parity is always odd). 

To diagonalize the Hamiltonian of Eq.\,\eqref{Hamilnok}, we propose the following (un-normalized) Bethe Ansatz state:
\begin{align}
\ket{\Psi}=&\prod_{\ell=1}^{Q-1}(U-\omega_\ell) \ket{r=1}=\sum_{\ell=0}^{Q-1}{\sf e}_{Q-\ell-1}U^\ell \ket{r=1}, \nonumber \\
&\langle s\ket{\Psi}=\frac{q_Q^s}{\sqrt{Q}}\prod_{\ell=1}^{Q-1}(\omega^{s-1}-\omega_\ell),
\end{align}
where 
${\sf e}$'s are elementary symmetric polynomials in $Q-1$ variables $\omega_\ell$:
\begin{align*} 
{\sf e}_0   &={\sf e}_Q=1,\quad {\sf e}_{1}=-\sum_{\ell=1}^{Q-1} \omega_\ell,\quad
{\sf e}_{2}=\sum_{\ell_1<\ell_2} \omega_{\ell_1}\omega_{\ell_2},\\
&\cdots \cdots, \quad {\sf e}_{Q-1}= {\sf e}_{-1}=(-1)^{Q-1}\omega_{1}\cdots \omega_{{Q-1}}.
\end{align*} 
The energy eigenvalues are given by 
\begin{eqnarray}
 \epsilon={\sf e}_1+{\sf e}_{Q-1}+(-1)^Q (\omega^* \, q_Q+ \omega \, q_Q^*) ,
\end{eqnarray}
with ${\sf e}$'s satisfying coupled polynomial equations
\begin{align*}
{\sf e}_{Q-\ell-1}\left\{\epsilon-(-1)^Q \left[\omega^\ell q_Q+(\omega^*)^\ell q_Q^*\right]\right\}
= {\sf e}_{Q-\ell}+{\sf e}_{Q-\ell-2},
\end{align*}
for $\ell=0,1,\cdots,Q-2$ and with $q_Q=e^{i \frac{\pi}{2 Q}[1+(-1)^Q]}$.

\section{The Bloch Ansatz vs. the Bethe Ansatz of Ref.\,[27]}
\label{compare}

Usually the Bethe Ansatz refers to an Ansatz for the exact wavefunctions of a one-dimensional interacting quantum model. It reduces the original problem with exponential complexity to a problem with polynomial complexity by solving coupled polynomial equations, i.e., the Bethe Ansatz equations of spectral parameters parameterizing the Ansatz wavefunctions. However, in the nineties, P. B. Wiegmann and A. V. Zabrodin\cite{Wiegmann} made a connection of the noninteracting Harper model to the quantum group $U_q(s\ell_2)$, and found that for the special case $\lambda(k_x,k_y)=0$, the spectrum can be written as $\epsilon=q^Q2\sin(-\pi \phi)\sum_{\ell=1}^{Q-1}\omega_\ell$, with $q=e^{i\pi \phi}$ and spectral parameters $\omega_\ell$ determined by Bethe Ansatz equations\cite{correction},
\begin{align}\label{BetheAnsatzEquations}
\frac{\omega_\ell^2+q}{q\omega_\ell^2+1}=q^Q\prod_{\ell^\prime=1}^{Q-1}\frac{q\omega_\ell-\omega_{\ell^\prime}}{\omega_\ell-q\omega_{\ell^\prime}}\quad(\ell=1,\dots,Q-1),
\end{align}
where the index ${\ell^\prime}=\ell$ should be included in the product and 
$\omega_{\ell^\prime}\neq\omega_\ell$ if ${\ell^\prime}\neq \ell$. 

For the noninteracting Harper model, both our Bloch Ansatz for the cylinder and the Bethe Ansatz of Ref.\,[\onlinecite{Wiegmann}] for the torus lead to a set of coupled polynomial equations as the usual Bethe Ansatz does. However, leaving aside the fact that one Ansatz can calculate boundary states and the other cannot, there are fundamental differences between the two. 
\begin{itemize}
	\item 
	For the cylinder, while the Bloch Ansatz Eq.\,(\ref{det(B)arbitraryBC}) reduces the complexity of diagonalizing a $(QN-\nu)\times (QN-\nu)$ matrix to that of diagonalizing a $Q\times Q$ and also a $(Q-1)\times(Q-1)$ matrices when $\nu=1$, it does not solve the diagonalization problem of the $Q\times Q$ and $(Q-1)\times(Q-1)$ matrices, which generally needs to be solved numerically.
	\item For the torus, while Bloch's theorem plays the same role as the Bloch Ansatz at the level of reducing complexity, Ref.\,[\onlinecite{Wiegmann}] does not reduce complexity of the problem but changes the methodology of diagonalizing a $Q\times Q$ matrix to that of solving $Q-1$ coupled nonlinear equations, and numerically solving the Bethe Ansatz Eq.\,(\ref{BetheAnsatzEquations}) solves the problem when $\lambda(k_x,k_y)=0$.
\end{itemize}

Furthermore, the spectral parameters $z_\ell$ of our paper and $\omega_\ell$ of Ref.\,[\onlinecite{Wiegmann}] can have completely different distributions on the complex plane as well. For example, for the torus geometry, $z_\ell$ takes values all from the unit circle as we mentioned, but this is not true for $\omega_\ell$. It is instructive to look more closely at two special cases, $\phi=1/3$ and $\phi=2/3$. Both have Eq.\,(\ref{bulkpolynomial}) as 
\begin{align}
	\epsilon^3-6\epsilon+2\lambda(k_x,k_y)=0.\nonumber
\end{align}
When $\lambda(k_x,k_y)=0$, the above equation is simplified to $\epsilon^3-6\epsilon=0$, which has three different energy eigenvalues $\{-\sqrt{6},0,\sqrt{6}\}$. Parallelly, for $\phi=1/3$, Eq.\,(\ref{BetheAnsatzEquations}) leads to $\{\omega_1=e^{i3\pi/4},\omega_2=e^{i5\pi/4}\}$ for $\epsilon=-\sqrt{6}$, $\{\omega_1=-1,\omega_2=1\}$ for $\epsilon=0$, and $\{\omega_1=e^{i\pi/4},\omega_2=e^{i7\pi/4}\}$ for $\epsilon=\sqrt{6}$. However, for $\phi=2/3$, solving Eq.\,(\ref{BetheAnsatzEquations}) leads to roots $\{\omega_1,\omega_2\}$ totally off the unit circle. Other than $\{-i,i\}$ corresponding to $\epsilon=0$, we have $\{-\sqrt{2-\sqrt{3}},\sqrt{2+\sqrt{3}}\}$ corresponding to $\epsilon=-\sqrt{6}$ and $\{-\sqrt{2+\sqrt{3}},\sqrt{2-\sqrt{3}}\}$ corresponding to $\epsilon=\sqrt{6}$. Furthermore, there are two additional roots with $\omega_1=\omega_2=1$ corresponding to $\epsilon=-2\sqrt{3}$ and $\omega_1=\omega_2=-1$ corresponding to $\epsilon=2\sqrt{3}$.

\section{Summary and conclusions}
\label{Summary}

In this paper, we studied analytically both bulk and boundary properties of the Harper model in various geometries\cite{explanation}. Based on a recent generalization of  Bloch's theorem\cite{PRB1}, we presented a Bloch Ansatz for the Harper Hamiltonian with an arbitrary rational magnetic flux $\phi=P/Q$ and cylindrical boundary conditions.
When the cylinder is of finite length $QN-1\,(N\in\mathds{Z}^+)$, we solved the associated Bloch Ansatz equations in complete analytic form. We found that the energy spectrum of boundary states has no dependence on the system size $N$, allowing us to construct a quasi-one-dimensional boundary Hamiltonian $H_\partial(k_y)$ that is exact and 
describes both two edges simultaneously. When the cylinder is of length $QN-\nu\,(\nu\neq 1)$, the boundary spectrum does depend on $N$. However, this dependence disappears in the limit $N\rightarrow \infty$. In this limit, we found that $H_\partial(k_y)$ can describe only half of the boundary states, while the other half is described by $H_\partial\big(k_y+2\pi\phi(1-\nu)\big)$.  The two boundary Hamiltonians coincide only in the case $\nu=1$. Our Bloch Ansatz puts boundary conditions into a perspective that facilitates boundary engineering and allows a better control of experimental imitations of the Harper model that, by construction, must include a boundary. 

Using arguments of gauge invariance, we proved that energies of bulk states, realizing the famous Hofstadter's butterfly, depend on a single geometry-dependent spectral parameter. This fact leads to a spectrum that shares the exact same functional form for the cylinder with $\nu=1$ and the torus with general twisted boundary conditions. Because of this functional form, bulk energies for a finite cylinder of length $QN-1$ correspond to exact energies for a finite torus with periodic and anti-periodic boundary conditions of length $QN$. 
Furthermore, we argued that the (boundary) bulk spectrum of a semi-infinite cylinder in 
an irrational magnetic field is (the complement of) a Cantor set. Finally, realizing that the bulk projection of 
the Harper Hamiltonian is a linear form over a deformed Weyl algebra, we introduced a Bethe Ansatz valid for both 
cylinder and torus geometries that differs from the one presented in Ref.\,[\onlinecite{Wiegmann}].

\section{Acknowledgments}   

G.O acknowledges support from the US Department of Energy grant  DE-SC0020343.

\appendix

\section{Gaussian dualities}\label{duality}

Can a superconductor display the same Cantor set structure as the Harper model? In this appendix, we briefly discuss the possibility of using duality transformations to address this question. 
After setting the stage, we describe a rather simple duality transformation that maps the Harper model to a strongly coupled superconductor (i.e., where the hopping terms vanish or are small). Roughly speaking, a duality transformation is any isometric (for the underlying Hilbert spaces) mapping (similarity transformation) of a strongly interacting model Hamiltonian into a weakly interacting one such that locality is preserved\cite{BADprl,GoodAP}. For quadratic fermionic Hamiltonians, this idea is realized by mappings, dubbed Gaussian dualities in Ref.\,[\onlinecite{FGD}], of particle non-conserving models, that is, superconductors, into particle conserving ones, that is, insulators or metals, at some specific filling. 

Consider the spinless Su-Schrieffer-Heeger \cite{SSH} (SSH) particle conserving Hamiltonian  and the Majorana chain of Kitaev \cite{Kitaev}. For periodic or open boundary conditions and an odd number of sites, there exists a Gaussian duality transformation that maps the SSH model to the Kitaev chain at vanishing chemical potential \cite{FGD}.  This duality illustrates the transmutation of topological classes when transformations in Fock space are allowed. There are many other examples of Gaussian dualities in Ref.\,[\onlinecite{FGD}] with similar characteristics and in higher space dimensions. What is critical in all those cases is the presence of the particle-hole symmetry as an internal symmetry. By contrast, the Harper model does not display any internal symmetries. 
Nonetheless, it is possible to map the Harper model to a local superconductor, and the key is to exploit the bipartite nature of the underlying square lattice. In the context of BCS theory, with spin as the source of the bipartition, the kind of duality we show next goes back to Anderson\cite{FGD}.

Suppose the mode indices of some Hamiltonian can be equally partitioned into $\x$ and $\y$ labels, and the Hamiltonian is of the form $\widehat{H}=\sum (t_{\x\y}c^\dagger_\x c_\y+ {\rm H.c.})- V \widehat{N}$, with \(\widehat{N}=\sum c^\dagger_\x c_\x+\sum c^\dagger_\y c_\y\) the number operator and \(V\) the chemical potential. The Harper model just represents a special case with $V=0$ and the mode indices $(m,n)$ in Eq.\,\eqref{Harper-Hofstadter}. Then, the local Gaussian transformation 
\(
c_\x\mapsto c_\x,\ c_\y\mapsto c_\y^\dagger 
\) 
maps the number conserving Hamiltonian to a dual, strongly interacting superconductor 
$
\widehat{H}^D=\sum (t_{\x\y}c^\dagger_\x c_\y^\dagger+{\rm H.c.})-V \widehat{N}^D ,
$
with 
\(
\widehat{N}^D=\sum c^\dagger_\x c_\x+\sum c_\y c_\y^\dagger
\)
a symmetry of the dual superconductor. Specializing to the Harper model, the dual superconductor features by construction surface modes ($Q>2$) and a Cantor set energy spectrum inherited from its dual Harper model.

\section{The Bloch Ansatz for $z_1=z_2=\pm 1$}
\label{degenerateroots}

For $z_1=z_2=\pm 1$ (associated with bulk solutions if they exist), the $2Q$ independent solutions for the bulk equation are
\begin{align}
\ket{\psi_s}&=\partial_{z_1}^{s-1}[\ket{z_1}\otimes|u(\epsilon,z_1)\rangle]\quad(s=1,2),\nonumber\\
\ket{\psi_s^+}&=\ket{N}\otimes|s+1\rangle\quad(s=1,2,\dots,Q-1),\nonumber\\
\ket{\psi_s^-}&=\ket{1}\otimes|s\rangle\quad(s=1,2,\dots,Q-1),\nonumber
\end{align}
A general Ansatz solution can be constructed as
\begin{align}\label{Bethe-ansatz2}
\hspace*{-0.1cm}\ket{\epsilon}=
\hspace*{-0.05cm}\sum_{s=1}^2\alpha_s\ket{\psi_s}+\hspace*{-0.05cm}\sum_{s=1}^{Q-1}\left(\alpha_{2+s}\ket{\psi_s^+}+\alpha_{Q+1+s}\ket{\psi_s^-}\right),\hspace*{-0.1cm}
\end{align}
which, similar as in Sec.\,\ref{BetheAnsatzSolutions}, leads to a system of linear equations, $B(\epsilon)[\alpha_1\,\alpha_2\,\cdots\,\alpha_{2Q}]^\text{T}=0$, but with
\begin{align}\label{det}
\text{det}&B(\epsilon)=(-1)^{Q-\nu+1}\bigg[\prod_{s=1}^\nu f_{Q-\nu+s}\bigg]\times\nonumber\\
&\text{det}\begin{bmatrix}
u_Q(\epsilon,z_1)				&\partial_{z_1}[u_Q(\epsilon,z_1)]\\
z_1^Nu_{Q-\nu+1}(\epsilon,z_1)	&\partial_{z_1}[z_1^Nu_{Q-\nu+1}(\epsilon,z_1)]
\end{bmatrix}.
\end{align}
As in the main text, let us focus on the case $\nu=1$. Then, the above equation simplifies to the following
\begin{align}
\text{det}B(\epsilon)
=(-1)^Qf_QNz_1^{N-1}u_Q^2(\epsilon,z_1),\nonumber
\end{align}
which admits a solution $u_Q(\epsilon,z_1)=0$ and therefore reduces to the discussion of Sec.\,\ref{boundarysolutions}, where $|\epsilon(r)\rangle$ of Eq.\,(\ref{edgestates}) will then correspond to bulk eigenstates if $z_1=\pm 1$ is satisfied at some special momenta $k_y$ and $|\epsilon(r)\rangle$ of Eq.\,(\ref{edgestates}) is just the Ansatz solution of Eq.\,(\ref{Bethe-ansatz2}) with the non-vanishing $\alpha_\ell$ being $\alpha_1=1$.

\section{The semi-infinite cylinder geometry}\label{semi-infinite}

In this appendix, we use the Bloch Ansatz to investigate the energy spectra, especially the bulk spectrum, of the Harper model in a semi-infinite cylinder with the termination at $x=1$. 
The Bloch Ansatz plays out differently in this geometry as compared to a finite cylinder. To guide our intuition, let us illustrate the main idea with a simple example first. 

Consider the simplest one-dimensional tight-binding model $\widehat{H}=-t\sum_{j=1}^\infty (c_j^\dagger c^{\;}_{j+1} + \text{H.c.})$, with the single-particle Hamiltonian $H=-t(T+T^\dagger)$ in terms of \(T=\sum_{j=1}^\infty |j\rangle \langle j+1|\), and the boundary projector \(P_\partial=|1\rangle\langle 1|\). Then, the Bloch Ansatz is 
\(
|\epsilon\rangle=\alpha_1|z\rangle +\alpha_2|z^{-1}\rangle 
\) when $z\neq\pm 1$ (or $|\epsilon\rangle=\alpha^\prime_1|z\rangle +\alpha^\prime_2\partial_z|z\rangle$ when $z=\pm 1$). Here, \(|z\rangle =\sum_{j=1}^\infty z^j|j\rangle\). 
The spectral parameter \(z\) is related to the energy \(\epsilon\)
by the equation $\epsilon=-t(z+z^{-1})$. Keeping this in mind, one can check that the boundary equation $P_\partial (H-\epsilon \mathds{1})|\epsilon\rangle=0$ becomes $t(\alpha_1+\alpha_2)|j=1\rangle=0$ (or $t\alpha^\prime_1|j=1\rangle=0$),
which fixes the coefficients of the Ansatz, \(\alpha_1=-\alpha_2=\alpha\) (or $\alpha^\prime_1=0$), but not the spectral parameter \(z\): 
any \(z\neq0\) specifies an algebraic solution of the difference equation 
\(H|\epsilon\rangle=\epsilon|\epsilon\rangle\) with corresponding \(\epsilon(z)\). However, when \(H\) is regarded as an operator on the Hilbert space of square summable sequences, its generalized eigenvectors, even if not square summable, must be bounded at infinity by a constant, which excludes both $|z|\neq 1$ and \(z=\pm 1\). 
One concludes that \(z=e^{i\phi}\) with \(\phi\in (-\pi,0)\cup(0,\pi)\), which leads to $\epsilon=-2t\cos\phi$ and $|\epsilon\rangle=2i\alpha\sum_{j=1}^\infty\sin(j\phi)|j\rangle$. As one would expect, this system does not display any boundary states. With the identification \(\phi \leftrightarrow k\), one can immediately check that the energy spectrum of the semi-infinite system coincides with that of the infinite one (Bloch bands) except that the former is missing energies at \(k=\{-\pi, 0\}\).

Now we are ready to tackle the Harper model in a semi-infinite cylinder. Since there is only one boundary, we need to truncate the Bloch Ansatz Eq.\,(\ref{Bethe-ansatz}) as follows,
\begin{align}
\ket{\epsilon}=
\sum_{\ell=1}^2\beta_\ell\ket{\psi_\ell}+\sum_{s=1}^{Q-1}\beta_{2+s}\ket{\psi_s^-},
\end{align}
with the definition of $\ket{z_\ell}$ of $\ket{\psi_\ell}=\ket{z_\ell}\otimes|u(\epsilon,z_\ell)\rangle$ changed to 
$\ket{z_\ell}=\sum_{j=1}^\infty z_\ell^j \ket{j}$ while the others remain. Similarly, we construct the boundary matrix
\begin{align}
&B(\epsilon)=
\begin{bmatrix}
-u_Q(\epsilon,z_1)&-u_Q(\epsilon,z_2)	&f_1	  &1	 &0		&\dots	&0\\
0		 &0		   						&1	  &f_2	 &1		&\ddots	&\vdots\\
0		 &0		   						&0	  &1     &\ddots&\ddots &0\\
\vdots	 &\vdots   						&\vdots&\ddots&\ddots&f_{Q-2}&1\\
0		 &0		   						&0	  &\dots &0		&1	   	&f_{Q-1}\\
0		 &0		   					 	&0	  &0	 &\dots	&0		&1
\end{bmatrix},\nonumber
\end{align}
which is a $Q\times(Q+1)$ matrix satisfying the linear equations $B(\epsilon)[\beta_1\,\beta_2\,\cdots\,\beta_{Q+1}]^\text{T}=0$. 
Obviously, $\beta_1$ and $\beta_2$ are the only possible non-vanishing coefficients of the Bloch Ansatz, satisfying the following relation
\begin{align}
u_Q(\epsilon,z_1)\beta_1+u_Q(\epsilon,z_2)\beta_2=0.
\end{align}

When $u_Q(\epsilon,z_1)u_Q(\epsilon,z_2)=0$, the discussion goes back to Sec.\,\ref{boundarysolutions}, and we get the same boundary states as Eq.\,(\ref{edgestates}), with boundary energy eigenvalues determined by 
Eq.\,(\ref{edgepolynomial}). However, note that Eq.\,(\ref{edgestates}) with $|z_1|>1$ are now associated with divergent semi-infinite sequences.

When $u_Q(\epsilon,z_1)u_Q(\epsilon,z_2)\neq 0$, we have the freedom to choose $u_Q(\epsilon,z_1)=u_Q(\epsilon,z_2)$ so that $\beta_1=-\beta_2$, and we get eigenstates as Eq.\,(\ref{bulkstates}), with $z_\ell$ undetermined. However, since Eq.\,(\ref{bulkstates}) with $|z_\ell|\neq 1$ are also associated with divergent sequences, the only choices are  $|z_\ell|=1$. These states have the same energy as, and are in natural correspondence with, bulk states with the exception of \(z_\ell=\pm 1\) (see Appendix\,\ref{degenerateroots} for details when \(z_\ell=\pm 1\), which are associated with Eq.\,(\ref{edgestates}) at some special momenta $k_y$). Hence, the bulk spectrum of a semi-infinite cylinder (one termination at $x=1$) coincides with that of an infinite one (no terminations) up to a finite number of spectral points.

\section{Twisted boundary conditions}
\label{TwistedBCs}

In order to model the twisted boundary conditions in the $x$ direction, we perform a corner modification of Eq.\,(\ref{W}) as follows
\begin{align*}
W=|N\rangle\langle 1|\otimes\left(-te^{i\Theta}h_1\right)+\text{H.c.},
\end{align*}
where $\Theta=(\pi) \, 0$ corresponds to the usual (anti-) periodic boundary condition. 
We can then construct the same Bloch Ansatz as Eq.\,(\ref{Bethe-ansatz}) but with only $\alpha_1$ 
and $\alpha_2$ non-vanishing. Similarly, we can construct a simplified $2\times 2$ boundary matrix 
$B(\epsilon)$ that satisfies $B(\epsilon)[\alpha_1\,\alpha_2]^\text{T}=0$, where 
\begin{align*}
&B(\epsilon)=
\begin{bmatrix}
(z_1^Ne^{-i\Theta}-1)u_Q(\epsilon,z_1)	&(z_2^Ne^{-i\Theta}-1)u_Q(\epsilon,z_2)\\
z_1(e^{i\Theta}-z_1^N)u_1(\epsilon,z_1)	&z_2(e^{i\Theta}-z_2^N)u_1(\epsilon,z_2)
\end{bmatrix}.
\end{align*}
Using the equality $u_Q(\epsilon,z_\ell)=-z_\ell[f_1u_1(\epsilon,z_\ell)+u_2(\epsilon,z_\ell)]$, 
we simplify the expression of $\text{det}B(\epsilon)$ as follows
\begin{align*}
\text{det}B(\epsilon)=&(z_1^N+z_2^N-2\cos\Theta)\times\\
&[u_1(\epsilon,z_1)u_2(\epsilon,z_2)-u_1(\epsilon,z_2)u_2(\epsilon,z_1)],
\end{align*}
where $z_1^N+z_2^N-2\cos\Theta=0$ determines all $N$ solutions for each energy band. 
After some algebraic manipulations, we get $z_1=e^{i(2\pi j+\Theta)/N}=z_2^{-1}$ ($j=0,1,\dots,N-1$).

\section{Boundary states for an arbitrary $\nu$}\label{nuneq1}

In Sec.\,\ref{Diagonalization}, we mentioned that Eq.\,(\ref{det(B)arbitraryBC}) does not have analytical solutions 
for a finite-size system when $\nu\neq 1$. In this appendix, we solve for boundary states analytically in the limit 
$N\rightarrow \infty$.

Let us start from Eq.\,(\ref{det(B)arbitraryBC}). A close examination shows that $u_Q(\epsilon,z_1)=0$ together 
with $|z_1|<1$ [or $u_Q(\epsilon,z_2)=0$ with $|z_2|<1$] is always a solution of Eq.\,(\ref{det(B)arbitraryBC}) 
whatever the value of $\nu$. From Sec.\,\ref{boundarysolutions}, we know that $u_Q(\epsilon,z_1)=0$ and 
$|z_1|<1$ are associated with boundary states localized at $x=1$. Thus, when $N\rightarrow \infty$, it is clear 
that boundary conditions at $x=L_x=QN-\nu$ will not influence boundary states localized at $x=1$. 

Similarly, $u_{Q-\nu+1}(\epsilon,z_1)=0$ together with $|z_1|>1$ [or $u_{Q-\nu+1}(\epsilon,z_2)=0$ with $|z_2|>1$] 
is another solution of Eq.\,(\ref{det(B)arbitraryBC}) that depends on the value of $\nu$. They are associated with 
boundary states localized at $x=L_x$. Boundary conditions at $x=L_x$ obviously  influence 
boundary states localized at $x=L_x$. 

\begin{figure}[t]
	\includegraphics[width=\columnwidth]{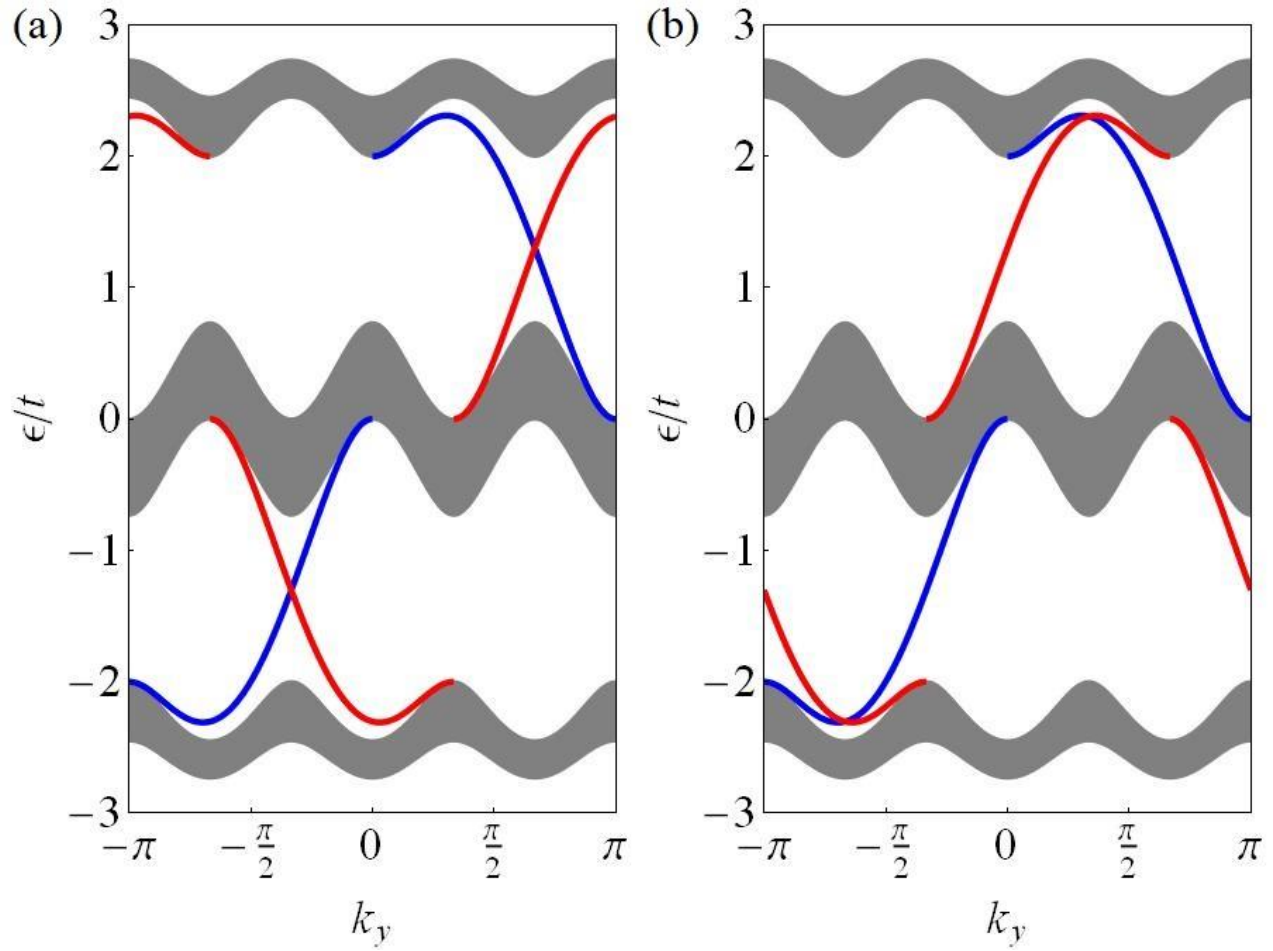}
	\caption{Band structures of the Harper model with $\phi=1/3$ in the cylinder for (a) the case $\nu=0$, and 
	(b) the case $\nu=2$. The blue lines are associated with boundary states localized at $x=1$, while the red lines 
	are associated with boundary states localized at $x=L_x$. Bulk bands are colored in gray.}
	\label{Fig_1/3}
\end{figure}

Just as Sec.\,\ref{boundarysolutions}, let us focus on $u_{Q-\nu+1}(\epsilon,z_1)=0$,  with $|z_1|>1$. In order to have a non-vanishing eigenvector $|u(\epsilon,z_1)\rangle=[u_1(\epsilon,z_1) \cdots u_Q(\epsilon,z_1)]^\text{T}$ of $H(z_1)$, 
one has  to satisfy the following relations
\begin{align}
\text{det}\left[M^{(k_y)}_{1-\nu,Q-1}\right]=0,\\ 
z_1=(-1)^{Q-1}\text{det}\left[M^{(k_y)}_{1-\nu,Q-2}\right],
\end{align}
which are generalizations of Eqs.\,(\ref{edgepolynomial}) and (\ref{z1}). Moreover, the above relations are equivalent to the following
\begin{align}
\text{det}\left[M^{[k_y+2\pi\phi(1-\nu)]}_{0,Q-1}\right]=0,\\
z_1=(-1)^{Q-1}\text{det}\left[M^{[k_y+2\pi\phi(1-\nu)]}_{0,Q-2}\right],
\end{align}
implying that, when compared with exact boundary solutions  for $\nu=1$ and in the limit $N\rightarrow \infty$, boundary states localized at $x=L_x$ for an arbitrary $\nu$ do not change but get shifted to $k_y+2\pi\phi(\nu-1)$. Thus, the total number of boundary states is a conserved quantity. For example, when $\phi=1/3$, Fig.\,\ref{Fig_1/3} displays the cases $\nu=0$ and $\nu=2$ that should be compared to $\nu=1$ plotted in Fig.\,\ref{Fig_p/q}\,(c).

\section{The Yang-Baxter equation}\label{YBE}

To make a connection to the Yang-Baxter equation, one can define the deformed Schwinger operators 
\begin{eqnarray}\hspace*{-0.5cm}
J_{\vec r,+} &=& (\omega^*)^{\frac{r_1 r_2}{2}} V_+(z)^{r_1} U^{r_2}, \nonumber \\
J_{\vec r,-} &=& \omega^{\frac{r_1 r_2}{2}} V_-(z)^{r_1} U^{r_2},
\label{Js}
\end{eqnarray}
where $\vec r=(r_1,r_2)$ with $r_{1,2}=0,1,\cdots,Q-1$. Moreover, their sets 
$\{J_{\vec r,+}\}, \{J_{\vec r,-}\}$ are closed under the Lie product
\begin{eqnarray}\hspace*{-0.5cm}
[ J_{\vec r,+}, J_{\vec s,+}]&=& 2 i \sin(\pi \phi \, \vec r \times \vec s) 
\, J_{\vec r + \vec s,+} , \nonumber \\
{[} J_{\vec r,-}, J_{\vec s,-}{]}&=& -2 i \sin(\pi \phi \, \vec r \times \vec s) 
\, J_{\vec r + \vec s,-} ,
\end{eqnarray}
with $\vec r \times \vec s=r_1 s_2 - s_1 r_2 $. Each set, excluding the element 
$J_{\vec 0,\pm}=\mathds{1}$, forms an $\mathfrak{su}(Q)$ algebra (they are the same algebra when $z^*=z^{-1}$). From their product
\begin{align}
J_{\vec r,\pm}J_{\vec s,\pm}= \sum_{\vec m, \vec n}(B_\pm)_{\vec r, \vec s}^{\quad\vec m, \vec n}J_{\vec m,\pm}J_{\vec n,\pm},
\end{align}
we obtain the commutation coefficients
\begin{align}
(B_\pm)_{\vec r, \vec s}^{\quad\vec m, \vec n}=\delta_{\vec r+\vec s}^{\vec m+\vec n}e^{\pm i\pi \phi(\vec r \times \vec s-\vec m \times \vec n)},
\end{align}
which satisfy the braid equation
\begin{align}
(\bold{B}_\pm)_{12}(\bold{B}_\pm)_{23}(\bold{B}_\pm)_{12}=(\bold{B}_\pm)_{23}(\bold{B}_\pm)_{12}(\bold{B}_\pm)_{23},
\end{align}
where $(\bold{B}_\pm)_{12}$ is the usual compact notation for $C\otimes D\otimes\mathds{1}$ if we define $B_\pm\equiv C\otimes D$, and $(\bold{B}_\pm)_{23}=\mathds{1}\otimes C\otimes D$, $(\bold{B}_\pm)_{13}=C\otimes\mathds{1}\otimes D$. Equivalently, we have the Yang-Baxter relation as follows
\begin{align}
(\bold{R}_\pm)_{23}(\bold{R}_\pm)_{13}(\bold{R}_\pm)_{12}=(\bold{R}_\pm)_{12}(\bold{R}_\pm)_{13}(\bold{R}_\pm)_{23},
\end{align}
with $(R_\pm)_{\vec s, \vec r}^{\quad\vec m, \vec n}\equiv(B_\pm)_{\vec r, \vec s}^{\quad\vec m, \vec n}$.


\begin{thebibliography}{}

\bibitem{Klitzing}
K. von Klitzing, G. Dorda, and M. Pepper, 
\textit{New Method for High-Accuracy Determination of the Fine-Structure Constant Based on Quantized Hall Resistance}, 
Phys. Rev. Lett. {\bf 45}, 494 (1980).

\bibitem{Chiu}
C.-K. Chiu, J. C. Y. Teo, A. P. Schnyder, and S. Ryu, 
\textit{Classification of topological quantum matter with symmetries}, 
Rev. Mod. Phys. {\bf 88}, 035005 (2016).

\bibitem{Harper}
P. G. Harper, 
\textit{Single band motion of conduction electrons in a uniform magnetic field}, 
Proc. Phys. Soc. A {\bf 68}, 874 (1955).

\bibitem{Azbel}
M. Ya. Azbel', 
\textit{Energy spectrum of a conduction electron in a magnetic field}, 
Soviet Phys. JETP {\bf 19}, 634 (1964).

\bibitem{Hofstadter}
D. R. Hofstadter, 
\textit{Energy levels and wave functions of Bloch electrons in rational and irrational magnetic fields}, 
Phys. Rev. B {\bf 14}, 2239 (1976).

\bibitem{TKNN}
D. J. Thouless, M. Kohmoto, M. P. Nightingale, and M. den Nijs, 
\textit{Quantized Hall Conductance in a Two-Dimensional Periodic Potential}, 
Phys. Rev. Lett. {\bf 49}, 405 (1982).

\bibitem{Hatsugai}
Y. Hatsugai, 
\textit{Chem Number and Edge States in the Integer Quantum Hall Effect}, 
Phys. Rev. Lett. {\bf 71}, 3697 (1993).

\bibitem{Ballentine}
L. E. Ballentine, \textit{Quantum Mechanics: A Modern Development} (World Scientific, Singapore, 2015), Sec.\,11.3.

\bibitem{Fradkin}
E. Fradkin, \textit{Field Theories of Condensed Matter Physics} (Cambridge University Press, Cambridge, 2013), Sec.\,12.2.

\bibitem{Kuhl}
U. Kuhl and H.-J. St\"ockmann, \textit{Microwave Realization of the Hofstadter Butterfly}, 
Phys. Rev. Lett. {\bf 80}, 3232 (1998). 

\bibitem{Albrecht}
C. Albrecht, J. H. Smet, K. von Klitzing, D. Weiss, V. Umansky, and H. Schweizer, 
\textit{Evidence of Hofstadter's Fractal Energy Spectrum in the Quantized Hall Conductance},
Phys. Rev. Lett. {\bf 86}, 147  (2001). 

\bibitem{Ponomarenko}
L. A. Ponomarenko, R. V. Gorbachev, G. L. Yu, D. C. Elias, R. Jalil, A. A. Patel, A. Mishchenko, A. S. Mayorov, C. R. Woods, J. R. Wallbank, M. Mucha-Kruczynski, B. A. Piot, M. Potemski, I. V. Grigorieva, K. S. Novoselov, F. Guinea, V. I. Fal’ko, and A. K. Geim,
\textit{Cloning of Dirac fermions in graphene superlattices},
Nature {\bf 497}, 594 (2013).

\bibitem{Dean}
C. R. Dean, L. Wang, P. Maher, C. Forsythe, F. Ghahari, Y. Gao, J. Katoch, M. Ishigami, P. Moon, M. Koshino, T. Taniguchi, K. Watanabe, K. L. Shepard, J. Hone, and P. Kim,
\textit{Hofstadter's butterfly and the fractal quantum Hall effect in moir\'e superlattices},
Nature {\bf 497}, 598 (2013).

\bibitem{Hunt}
B. Hunt, J. D. Sanchez-Yamagishi, A. F. Young, M. Yankowitz, B. J. LeRoy, K. Watanabe, T. Taniguchi, P. Moon, M. Koshino, P. Jarillo-Herrero, and R. C. Ashoori,
\textit{Massive Dirac fermions and Hofstadter butterfly in a van der Waals heterostructure},
Science {\bf 340}, 1427 (2013).

\bibitem{Satija}
I. I. Satija, \textit{Butterfly in the Quantum World}, (Morgan \& Claypool Publishers, San Rafael, 2016). 

\bibitem{Bloch}
M. Aidelsburger, M. Atala, M. Lohse, J. T. Barreiro, B. Paredes, and I. Bloch, 
\textit{Realization of the Hofstadter Hamiltonian with Ultracold Atoms in Optical Lattices}, 
Phys. Rev. Lett. {\bf 111}, 185301 (2013).

\bibitem{Ketterle}
H. Miyake, G. A. Siviloglou, C. J. Kennedy, W. C. Burton, and W. Ketterle, 
\textit{Realizing the Harper Hamiltonian with Laser-Assisted Tunneling in Optical Lattices}, 
Phys. Rev. Lett. {\bf 111}, 185302 (2013).

\bibitem{Zoller}
M. Mancini, G. Pagano, G. Cappellini, L. Livi, M. Rider, J. Catani, C. Sias, P. Zoller, M. Inguscio, M. Dalmonte, and L. Fallani,
\textit{Observation of chiral edge states with neutral fermions in synthetic Hall ribbons}, 
Science {\bf 349}, 1510 (2015).

\bibitem{Spielman}
B. K. Stuhl, H.-I. Lu, L. M. Aycock, D. Genkina, and I. B. Spielman, 
\textit{Visualizing edge states with an atomic Bose gas in the quantum Hall regime},
Science {\bf 349}, 1514 (2015).

\bibitem{Comment}
In Table\,I of Ref.\,[\onlinecite{PRB2}], we gave a short positive answer to the first question, which will be explained in Sec.\,\ref{BetheAnsatzSolutions} of this paper in detail.

\bibitem{PRL1}
A. Alase,  E. Cobanera, G. Ortiz, and L. Viola,
\textit{Exact Solution of Quadratic Fermionic Hamiltonians for Arbitrary Boundary Conditions}, 
Phys. Rev. Lett. {\bf 117}, 076804 (2016).

\bibitem{BlochAnsatz}
E. Cobanera, A. Alase, G. Ortiz, and L. Viola, 
\textit{Exact solution of corner-modified banded block-Toeplitz eigensystems},
J. Phys. A \textbf{50}, 195204 (2017).

\bibitem{PRB1}
A. Alase,  E. Cobanera, G. Ortiz, and L. Viola,
\textit{Generalization of Bloch's theorem for arbitrary boundary conditions: Theory}, 
Phys. Rev. B {\bf 96}, 195133 (2017).

\bibitem{PRB2}
E. Cobanera, A. Alase, G. Ortiz, and L. Viola,
\textit{Generalization of Bloch's theorem for arbitrary boundary conditions: Interfaces and topological surface band structure},
Phys. Rev. B \textbf{98}, 245423 (2018).

\bibitem{FermionBoson}
Note that for particle-conserving noninteracting bosons, they share the same single-particle energy spectra as their fermionic counterparts, although the many-body spectra are very different because of Bose-Einstein condensation\cite{Qiaoru2020}.

\bibitem{Math}
A. Avila and S. Jitomirskaya,
\textit{The Ten Martini Problem}
Ann. Math. {\bf 170}, 303 (2009).

\bibitem{Wiegmann}
P. B. Wiegmann and A. V. Zabrodin,
\textit{Bethe-Ansatz for the Bloch Electron in Magnetic Field}, 
Phys. Rev. Lett. {\bf 72}, 1890 (1994).

\bibitem{Thouless}
D. J. Thouless,
\textit{Bandwidths for a quasiperiodic tight-binding model}, 
Phys. Rev. B {\bf 28}, 4272 (1983).

\bibitem{Santos}
J. Wang and L. H. Santos,
\textit{Classification of Topological Phase Transitions and van Hove Singularity Steering Mechanism in Graphene Superlattices}, 
Phys. Rev. Lett. {\bf 125}, 236805 (2020).

\bibitem{HatsugaiPRB}
Y. Hatsugai, 
\textit{Edge states in the integer quantum Hall effect and the Riemann surface of the Bloch function},
Phys. Rev. B {\bf 48}, 11851 (1993).

\bibitem{Ryu}
S. Ryu, A. P. Schnyder, A. Furusaki, and A. W. W. Ludwig, 
\textit{Topological insulators and superconductors: tenfold way and dimensional hierarchy}, 
New J. Phys. {\bf 12}, 065010 (2010).

\bibitem{Qiaoru2020}
Q.-R. Xu, V. P. Flynn, A. Alase, E. Cobanera, L. Viola, and G. Ortiz, 
\textit{Squaring the fermion: The threefold way and the fate of zero modes}, 
Phys. Rev. B {\bf 102}, 125127 (2020).

\bibitem{Laughlin}
R. B. Laughlin, 
\textit{Quantized Hall conductivity in two dimensions},
Phys. Rev. B {\bf 23}, 5632 (1981).

\bibitem{Halperin}
B. I. Halperin, 
\textit{Quantized Hall conductance, current-carrying edge states, and the existence of extended states in a two-dimensional disordered potential},
Phys. Rev. B {\bf 25}, 2185 (1982).

\bibitem{Schwinger} 
J. Schwinger, {\it Quantum Mechanics: Symbolism of Atomic Measurements}, Springer-Verlag, Berlin (2001).

\bibitem{Ortiz2012} 
G. Ortiz, E. Cobanera, and Z. Nussinov, 
\textit{Dualities and the phase diagram of the $p$-clock model}, 
Nucl. Phys. B {\bf 854}, 780 (2012).

\bibitem{correction}
The Bethe Ansatz equations in Ref.\,[\onlinecite{Wiegmann}] are different from Eq.\,(\ref{BetheAnsatzEquations}) and read as $\frac{\omega_\ell^2+q}{q\omega_\ell^2+1}=q^Q\prod_{{\ell^\prime}=1,{\ell^\prime}\neq\ell}^{Q-1}\frac{q\omega_\ell-\omega_{\ell^\prime}}{\omega_\ell-q\omega_{\ell^\prime}}$.

\bibitem{explanation}
For simplicity, in this work we focused on various geometries for the square lattice even though other lattices are just as tractable with minor changes of procedure. For instance, one should take care of the distribution of fluxes in the lattice (Peierls substitution) and write down the associated model Hamiltonian. Depending upon the lattice, one should modify the basis for the bulk equation and, therefore, the Bloch Ansatz for the boundary equation. Finally, one needs to derive the corresponding boundary matrix $B(\epsilon)$ from the boundary equation and find solutions of the associated equation $\text{det}B(\epsilon)=0$.

\bibitem{BADprl}
E. Cobanera, G. Ortiz, and Z. Nussinov,
\textit{A Unified Approach to Quantum and Classical Dualities},
Phys. Rev. Lett. \textbf{104}, 020402 (2010).

\bibitem{GoodAP}
E. Cobanera, G. Ortiz, and Z. Nussinov,
\textit{The bond-algebraic approach to dualities},
Adv. Phys. \textbf{60}, 679 (2011).

\bibitem{FGD}
E. Cobanera and G. Ortiz, 
\textit{Equivalence of topological insulators and superconductors},
Phys. Rev. B \textbf{92}, 155125 (2015).

\bibitem{SSH}
W. P. Su, J. R. Schrieffer, and A. J. Heeger, 
\textit{Solitons in Polyacetylene},
Phys. Rev. Lett. {\bf 42}, 1698 (1979).

\bibitem{Kitaev}
A. Y. Kitaev, 
\textit{Unpaired Majorana fermions in quantum wires},
Phys.-Usp. {\bf 44}, 131 (2001).
	
\end{thebibliography}
\end{document}